\documentclass[aps,prd,twocolumn,preprintnumbers,superscriptaddress,nofootinbib,showpacs]{revtex4-2}
\usepackage{aas_macros}
\usepackage{graphicx}
\usepackage{ulem}
\usepackage{comment}
\usepackage{subcaption}
\usepackage{bm,latexsym,amsmath,amssymb,amsfonts,mathrsfs}
\usepackage{color}
\usepackage[colorlinks=true,linkcolor=magenta,citecolor=magenta,urlcolor=magenta]{hyperref}
\usepackage{hyperref}
\usepackage{orcidlink}
\usepackage{tikz}
\usepackage{caption}
\usetikzlibrary{arrows.meta,positioning}
\newcommand*{\D}{{\rm d}}

\newcommand{\simgt}{\lower.5ex\hbox{$\; \buildrel > \over \sim \;$}}
\newcommand{\simlt}{\lower.5ex\hbox{$\; \buildrel < \over \sim \;$}}

\newcommand{\dd}{\mathop{}\!\mathrm{d}}

\begin{document}

\title{Analytic and numerical study of dark matter halos in the Horndeski theory} 

\author{Kohei Onoda}
\affiliation{Department of Physics, Nagoya University, Nagoya 464-8602, Japan}

\author{Shun Arai\orcidlink{0000-0002-2527-3705}}
\email{s-arai@daiichi-koudai.ac.jp}
\affiliation{Department of Information, Artificial Intelligence and Data Science,
Daiichi Institute of Technology, Kagoshima 890-0052, Japan}
\affiliation{Kobayashi-Maskawa Institute for the Origin of Particles and the Universe (KMI), Nagoya University, Nagoya 464-8602, Japan}

\author{Daisuke Yamauchi}
\affiliation{Department of Physics, Faculty of Science, Okayama University of Science, 1-1 Ridaicho, Okayama, 700-0005, Japan}

\author{Shuichiro Yokoyama\orcidlink{0000-0002-1811-5477}}
\affiliation{Department of Physics, Nagoya University, Nagoya 464-8602, Japan}
\affiliation{Kobayashi-Maskawa Institute for the Origin of Particles and the Universe (KMI), Nagoya University, Nagoya 464-8602, Japan} 
\affiliation{Kavli Institute for the Physics and Mathematics of the Universe (WPI), UTIAS, The University of Tokyo, Kashiwa, Chiba 277-8583, Japan}

\author{Shogo Masaki}
\affiliation{Department of Information Engineering and Institute for Advanced Studies in Artificial
Intelligence, Chukyo University, Toyota, Aichi 470-0393, Japan}
\affiliation{Department of Physics, Nagoya University, Nagoya 464-8602, Japan}

\author{Kenji Hasegawa}
\affiliation{National Institute of Technology, Suzuka College, Shirokocho, Suzuka, Mie 510-0294, Japan}

\date{\today}

\begin{abstract}
The models of modified gravity that can explain the late-time accelerated expansion of the Universe must pass the local tests of gravity, introducing a screening mechanism so that any additional fifth force in gravity is suppressed at stellar scales. Dark matter halos, in contrast, by their extendedness in scale, may exhibit a transition of screening to small scales, providing a probe of modified gravity and the origin of the late-time acceleration. 
We perform comprehensive studies of dark matter halos with Vainshtein screening in the Horndeski theory.
Given that dark matter is cold and a dark matter halo is spherically symmetric, we analytically clarify the conditions for the halo profile under which the Vainshtein screening works effectively.
We find that the Vainshtein screening is less effective even within the Vainshtein radius as the profile approaches a core-like structure around the center of a halo, which motivates us to postulate that some features owing to the fifth force can survive during the dynamical evolution of a halo.  We numerically examine our postulate by a $N$-body simulation for an isolated halo with the assumption of Newtonian two-body interactions with the spherical external fifth force.
We make use of three proxies of the mass through hydrostatic dynamics, the Jeans equation, and gravitational lensing, all of which must be identical
in the case of Newtonian gravity as a weak-field approximation to General Relativity.
We find that these proxies can be biased relative to each other at a (quasi) equilibrium, quantitatively indicating that the fiducial model of gravity is not Newtonian. 
Our result affirmatively suggests that multiple mass proxies from observations are capable of testing gravity on halo scales.
\end{abstract}

\maketitle

\section{Introduction}
\label{sec:intro}
Modified gravity (MG) that extends Einstein's general relativity (GR) with extra degrees of freedom in the gravity sector can provide an alternative explanation for the late-time acceleration of the Universe against the conventional cosmological constant in the $\Lambda$-Cold-Dark-Matter ($\Lambda$CDM) model (cf. the latest reviews are in~\cite{Ishak:2018his, Kobayashi2019, Arai2023}). Provided that a cosmological model in MG drives the late-time acceleration, such a model must pass not only the cosmological observations but also the local tests of gravity. 
At present, GR is consistent with various local tests in the solar system, at least including the strong gravity regime probed by Hulse-Taylor pulsars and gravitational waves. To evade the constraints of the local tests, screening mechanisms are essential for MG, which enable the suppression of any fifth force in gravity arising from the extra degrees of freedom compared to GR at small scales. 

The Horndeski theory~\cite{Horndeski1974, Deffayet2009, Kobayashi2011} is one of the well-established frameworks in which the Vainshtein screening~\cite{Vainshtein1972, Kimura2012} is built in, as well as introducing the dynamics of a scalar field on cosmological scales that can explain the late-time acceleration (see the latest review in~\cite{Kobayashi2019}). In fact, the radius of Vainshtein screening, so-called the Vainshtein radius, for the solar system is about $100$ pc, and thus the fifth force is negligible compared to the Newtonian force in two-body interactions between stellar bodies. Consequently, few signatures of the fifth force can be observed in star clusters in which two-body relaxation is dominant. 

The dark matter halos, on the other hand, lie on intermediate scales over which the fifth force is still effective up to cosmological scales. Since a typical Vainshtein radius estimated from the enclosed mass of a halo is of the same order of magnitude as its virial radius, it is expected that there is a transient signature between the screened and the unscreened regions in a halo. Many studies have been carried out aiming to find any imprints of the fifth force in a dark matter halo both analytically and numerically. For instance, analytical studies predict deformation in the convergence field of the gravitational lensing~\cite{Narikawa2013}, given fixed density profiles. Some studies carry out cosmological $N$-body simulations for a specific model in the Horndeski theory, i.e., nDGP model~\cite{Lue:2005ya}, quantitatively showing a transient signature in higher-order cumulants of the density fluctuation~\cite{Hellwing:2017pmj}, or in an effective gravitational coupling $G_{\rm eff}(z,\boldsymbol{k})$~\cite{2009PhRvD..80j4006S, Brando:2023fzu} as a function of the redshift $z$ and the wavenumber $\boldsymbol{k}$ in Fourier space. Whether the signatures found above are generic among the models of Vainshtein screening in the Horndeski theory or not is still an open question. The difficulty is that the full $N$-body simulation for the Horndeski theory is numerically costly to solve the nonlinear scalar-field equation that consists of higher spatial derivatives~\cite{Barreira2013, Barreira2015} (see also the latest prescriptions in \citep{Brando:2023fzu}).

In this paper, we aim to clarify how dark matter haloes are affected by the Vainshtein screening in the Horndeski theory, keeping the generic parameterization of the Horndeski theory that describes a spherically symmetric halo. We begin with the previous study~\cite{Narikawa2013} (hereafter we shall call this NKYS13) in which the dark matter halo under the Vainshtein screening in the Horndeski theory is analytically investigated. NKYS13 reports oscillatory features in the convergence field of gravitational lensing, given a fixed density profile of the (generalized) Navarro-Frenk-White (NFW) profile~\citep{Navarro1997}. We point out that the assumption of the density profile should be relaxed, as the density profile is determined throughout the dynamical evolution of a halo. We shall analytically quantify how effective the Vainshtein screening is by changing the shape of the density profile. 

We shall also quantitatively analyze the dynamical evolution of a dark matter halo with $N$-body simulations.  We assume that (1) the two-body interaction between the dark matter particles is Newtonian and (2) the fifth force acts as an external central force. Since we are interested in the late-time acceleration, we focus on haloes that reach dynamical (quasi) equilibrium with the Newtonian gravity in the early stage of structure formation. This assumption is well aligned with the assumption that the fifth force turns on only at late time. For simplicity, we consider an isolated halo. We numerically provide the initial conditions for a halo, following~\cite{Ogiya:2017hbr}.

We shall employ multiple mass proxies that can trace the dynamics of halo formation ~\cite{1976A&A....49..137C, KomatsuSeljak2001, 1915MNRAS..76...70J, Tyson1990, KaiserSquires1993}. We make use of three proxies of the mass of a halo: the hydrodynamical mass, the Jeans mass from the Jeans equation, and the lensing mass, given that the Newtonian gravity is assumed to be fiducial. Crucially, the three mass proxies can deviate from each other if the gravity is not Newtonian in the noiseless limit. We shall compute the mass proxies given the model of a halo with Vainshtein screening, showing a bias among the mass proxies that can indicate a signature of the unscreened fifth force in the Horndeski theory.

The rest of the paper is written as follows. In Sec.~\ref{sec:vainshtein_in_horndeski_spherical_halo}, we summarize the Vainshtein screening in the Horndeski theory for a spherically symmetric system by following~\cite{Narikawa2013}, and we clarify how effective the Vainshtein screening is with respect to the dependence on the strong coupling energy scale $\Lambda$ and halo concentration $c_s$. In Sec.~\ref{sec:nbody_sim_spherical_haloes_Horndeski}, we examine how effective the fifth force can be within the Vainshtein radius, employing $N$-body simulations. In Sec.~\ref{sec:mass_proxies}, we introduce multiple mass proxies that differ in tracing the enclosed mass of the dark matter halo, showing distinguishable features of the fifth force in gravity. 
In Sec.~\ref{sec:conclusion}, we discuss how valid our results are compared to the previous studies,
and present conclusions of this paper.
Throughout the paper, we set the speed of light $c$ to unity, i.e., $c=1$, and the signature of the spacetime metric follows the convention $(-,+,+,+)$.

\section{Vainshtein screening with spherical halo model}
\label{sec:vainshtein_in_horndeski_spherical_halo}

We introduce a stationary spherical halo model in the Horndeski theory~\cite{Horndeski1974, Deffayet2009, Kobayashi2011} to introduce the Vainshtein screening mechanism~\cite{Vainshtein1972, Kimura2012} for a dark matter halo, following the derivation in Ref.~\cite{Narikawa2013}. To make the presentation simple, we shall begin with the field equations for a halo that are derived in~\cite{Narikawa2013}. As a concise reference, we describe the quantities that are computed from the arbitrary functions in the Horndeski theory in App.~\ref{app:Horndeski_halo_model_detailed}.

\subsection{Asymptotic solutions in Horndeski theory}
\label{ssec:asymptotic_solution}

Let us suppose that the late-time acceleration is driven by a scalar field $\phi(t,\boldsymbol{x})$ that covariantly couples to gravity in the Horndeski theory. 
Given spherical symmetry while allowing the time variation of the scalar field, the solution of the Vainshtein screening is obtained in Ref.~\cite{Kimura2012}. In Ref.~\cite{Kimura2012}, it is shown that the time variation of the gravitational coupling remains in the weak-field limit. Such time variation is strongly constrained by observing the time variation of the Newton constant~\cite{2004PhRvL..93z1101W, 2018CQGra..35c5015H}. Given the constraint, the scalar field is static during the dynamical timescale of halo evolution.
Then, the metric and the scalar field are defined as
\begin{align}
{\rm d}s^2 &= - (1 + 2\Phi(r)){\rm d}t^2 + (1-2\Psi(r))\delta_{ij}{\rm d}x^i {\rm d}x^j\,,\cr
\phi &= \phi_0 + \varphi(r)\,.
\end{align}
where $r$ is the radius from the halo center, $\Phi(r)$ is the Newton potential, and $\Psi(r)$ is the potential of the spatial curvature, respectively. Note that we assume that the gravitational potentials are also static. 

We describe the field equations derived in~\cite{Narikawa2013} as follows. By introducing the dimensionless variables as
\begin{align}
x(r) := \frac{1}{\Lambda^3}\frac{\varphi'}{r},\,\ \ 
A(r) := \frac{1}{M_{\rm Pl}\Lambda^3}
\frac{M(<r)}{8\pi r^3},\label{eq:def_A(r)}
\end{align}
where $\Lambda$, $M_{\rm Pl}$, and $M(<r):=4\pi\int_0^r\bar r^2\rho (\bar r){\rm d}\bar r$ respectively define the strong coupling scale of the scalar field, Planck mass, and enclosed mass. The gravitational field equations are given by
\begin{align}
\frac{M_{\rm Pl}}{\Lambda^3}\frac{\Phi'}{r}
&=
-\xi x+\beta x^3+A(r),
\label{eq:Phi-eq}
\\
\frac{M_{\rm Pl}}{\Lambda^3}\frac{\Psi'}{r}
&=
\xi x+\alpha x^2+\beta x^3+A(r),
\label{eq:Psi-eq}
\end{align}
and the scalar-field equation is given by
\begin{align}
P(x,A)
:=&
\xi A
+\left(\frac{\eta}{2}+3\xi^2\right)x
+\left(\mu+6\alpha\xi-3\beta A\right)x^2
\nonumber\\
&+\left(\nu+2\alpha^2+4\beta\xi\right)x^3
-3\beta^2x^5
=0.
\label{eq:varphi-eq}
\end{align}
Note that the prime denotes the derivative with respect to $r$. The six dimensionless parameters $\xi,\eta, \mu,\nu, \alpha$, and $\beta$ are determined by the functions of the Horndeski theory as shown in  Eq.~\eqref{eq:parametrize}. Eqs.~\eqref{eq:Phi-eq},~\eqref{eq:Psi-eq}, and~\eqref{eq:varphi-eq} are solved in the following way. Equation~\eqref{eq:varphi-eq} gives algebraic solutions for $x(r) = x[A(r)]$. By substituting $x(r)$ into Eq.~\eqref{eq:Phi-eq} or Eq.~\eqref{eq:Psi-eq}, the integration over $r$ gives $\Phi(r)$ and $\Psi(r)$, respectively. 

Eq.~\eqref{eq:varphi-eq} exhibits multiple solutions as the roots of $x(r)$. Let us divide the vicinity of the source: $A(r) \gg 1$ and the region far from the source: $A(r) \ll 1$. We select a branch that can recover Newtonian gravity at $A(r) \gg 1$ in the vicinity of the source, where the Vainshtein screening operates. We introduce the Vainshtein radius as the solution of the condition
\begin{align}
A(r_{\rm V}) = 1\,.
\end{align}

In the subsequent analysis, we assume $\xi > 0$ for simplicity.
In the region outside the Vainshtein radius, $A(r) \ll 1$, we neglect the nonlinear terms in Eq.~\eqref{eq:varphi-eq}, and there is always a decaying solution as
\begin{align}
    x \simeq x_{\rm out} = -\frac{2\xi A}{\eta + 6\xi^2}\,.
    \label{eq:outer_solution_beta_neq_0}
\end{align}
Note that $\eta + 6\xi^2 > 0$ must be satisfied so that the kinetic term of the scalar field has the right sign.

We then look for inner solutions with $x\gg 1$ for $A\gg 1$, because scalar excitation is suppressed due to nonlinearity so we may expect that GR is reproduced.
To do this, we construct a solution by considering the balance of the terms with the large contributions in Eq.~\eqref{eq:varphi-eq} under $x\gg 1$ and $A\gg 1$.
Given Eq.~\eqref{eq:varphi-eq}, there are distinguished inner solutions depending on whether $\beta$ is zero or nonzero. This is because the term $\beta A$ vanishes when $\beta =0$. Otherwise, it is principal in Eq.~\eqref{eq:varphi-eq} when $A(r) \gg 1$. Let us consider the following two cases with the dependence of the value of $\beta$.\\
(i) $\beta \neq 0$. Since standard gravity is not reproduced in $\beta <0$, we will not consider this parameter region hereafter. When $\beta >0$, the one stable constant inner solution lies in
\begin{align}
    x\simeq x_{\rm in} \in (x_-, 0)\,,\ \ \ x_{-} = -\sqrt{\frac{\xi}{3\beta}}\,,    \label{eq:inner_solution_beta_neq_0}
\end{align}
as $P(0,A) = \xi A > 0$ and $P(x_-,A) < 0$ are satisfied that guarantee the intermediate value theorem. 
Note that the solution approaches to $x = x_-$ in the limit $A(r) \rightarrow \infty$. 
Given a finite value of $A(r)$, we numerically find the solution in the domain in Eq.~\eqref{eq:inner_solution_beta_neq_0}.
Since this branch reproduces the correct behavior of Newtonian gravity, 
this is the Vainshtein solution we are looking for when $\beta >0$.\\ 
(ii) $\beta = 0$.
The inner solution is given by
\begin{align}
    x^3 \simeq x^3_{\rm in} = -\frac{\xi A}{\nu + 2\alpha^2}\,,
    \label{eq:inner_cubic}
\end{align}
where we have assumed $\nu +2\alpha^2\neq 0$.
This solution can be smoothly connected to the decaying outer solution Eq.~\eqref{eq:outer_solution_beta_neq_0} only for $\nu +2\alpha^2 >0$ when the stability conditions of the inner solution is satisfied (see Ref.~\cite{Narikawa2013}).
Hereafter, we separately investigate the two cases: (i) $\beta \neq 0$ and (ii) $\beta = 0$.

\subsection{Scaling relations for fifth-force acceleration }
\label{ssec:scaling_of_Delta_acc}

Let us investigate the Vainshtein screening for the spherical halo
varying the spectral shapes of the matter density profile of the halo. Varying the density profile deforms the power of $A(r)$ with $r$ across the various radii, and thus we expect that the fifth force that $x(r)$ mediates also varies in scale. In what follows, we aim to derive a criterion that characterizes how effective the Vainshtein screening is, even within the Vainshtein radius, given various types of the density profile.

Using Eq.~\eqref{eq:Phi-eq}, the term proportional to $A(r)$ in the right-hand side corresponds to the Newtonian contribution, while the remaining terms represent modifications due to the scalar field. We define the fractional deviation of the inward acceleration $-\Phi'/r$ as
\begin{align}
\Delta_{\rm acc}(r)
:=
\frac{-\xi x(r)+\beta x^3(r)}{A(r)}\,.
\label{eq:Delta_acc_def}
\end{align}
For non-relativistic test particles, $\Delta_{\rm acc}$ can be interpreted as a deviation of the effective gravitational constant:
\begin{align}
G_{\rm eff}
=
(1+\Delta_{\rm acc})\,G_{\rm N}\,,
\label{eq:Delta_acc_G_eff}
\end{align}
where we have defined the Newton constant as $G_{\rm N}=1/8\pi M_{\rm Pl}^2$\,.
Thus, $\Delta_{\rm acc}$ can be observed as an extra acceleration compared to Newtonian gravity. In the following subsections, we shall derive the scaling relations of $\Delta_{\rm acc}$ with $r$, and construct a quantitative criterion for the Vainshtein screening conditioned by the parameters of the halo density profile.

To examine the scaling behavior in the halo, we assume that the density profile simply behaves as
\begin{align}
\rho
(r) \propto r^{-\gamma},
\qquad
0 \le \gamma < 3.
\end{align}
Then the enclosed mass scales as
\begin{align}
M(<r) \propto r^{3-\gamma},
\end{align}
and since $A(r) \propto M(<r)/r^3$, we obtain
\begin{align}
A(r) \propto r^{-\gamma}.
\label{eq:A_to_gamma}
\end{align}
Note that $A(r) \rightarrow \infty$ as $r \rightarrow 0$, satisfying the condition $A(r) \gg 1$ if $\gamma > 0$. Otherwise, $A(r)$ is capped by a finite value.

By making use of the asymptotic solutions for $x$ obtained in the previous subsection, we can obtain the fractional deviation of the inward acceleration, $\Delta_{\rm acc}$, as
\begin{align}
\Delta_{\rm acc}(r)
=
\frac{2\xi^2}{\eta + 6\xi^2},
\label{eq:Delta_outer}
\end{align}
in the outer region from Eq.~\eqref{eq:outer_solution_beta_neq_0}, and
\begin{align}
\Delta_{\rm acc}(r)
\propto
\begin{cases}
\displaystyle
\frac{1}{A(r)}
\propto
r^{\gamma} &~(\beta>0), 
\\[1ex]
\displaystyle
\frac{|x|}{A}
\propto
A^{-2/3}
\propto
r^{\frac{2}{3}\gamma} & ~(\beta = 0).
\end{cases}
\,
\label{eq:Delta_scaling_beta_neq_0}
\end{align}
in the inner region from Eq.~\eqref{eq:inner_solution_beta_neq_0} for $\beta>0$
and Eq.~\eqref{eq:inner_cubic} for $\beta = 0$, respectively.

From the above expressions, we find that the acceleration deviation is constant with respect to radius in the outer region. Recalling that $\eta + 6\xi^2 > 0$, the extra acceleration always makes a constant enhancement of gravitational coupling.
For instance, if $\eta\simeq 0$ or $\xi^2 \gg \eta$, we have
$\Delta_{\rm acc} \simeq 1/3,$ corresponding to approximately $30\%$ enhancement of gravity. In the inner region, on the other hand, the above expressions imply that for shallow central density profiles (small $\gamma$), the suppression of $\Delta_{\rm acc}$ can be weakened near the center of a halo.
In particular, for cored profiles ($\gamma\simeq 0$),
the suppression does not strengthen toward $r\to 0$, and the Vainshtein screening may saturate. This feature is not apparent when considering only a point-mass source ($A \propto r^{-3}$), and it is an important characteristic for extended mass distributions such as halos. Consequently, as we cannot use the inner solution for the cored profiles, we shall examine this case in more detail in the next subsection.

\subsection{Fifth force in cored profiles and unscreened residual}
\label{ssec:core_profile_analysis}

As we have seen so far, the behavior of Vainshtein screening is expected to differ significantly from that of a point source when we consider extended mass distributions.
In particular, based on the scaling behavior discussed above, for the cored profiles it is conceivable that the screening mechanism might not function even near the center of the halo.
Then, we revisit the behavior of $A(r)$ and the strength of screening in a halo with a finite-density core. 

First, let us consider the simplest case with the following core profile of the local density:
\begin{align}
\rho(r)
\simeq
\rho_{\rm core} = {\rm const.}
\qquad (r \ll r_{\rm core}).
\label{eq:constant_core}
\end{align}
Then the enclosed mass is derived as
\begin{align}
M(<r)
\simeq
\frac{4\pi}{3}
\rho_{\rm core}
r^3,
\end{align}
so that
\begin{align}
A(r)
\simeq
A_{\rm core}
:=
\frac{1}{2L^2}
\frac{\rho_{\rm core}}{\rho_{\rm crit}},
\label{eq:A_core_constant}
\end{align}
where we have introduced a dimensionless scaling parameter $L$ that determines the characteristic energy scale $\Lambda$ by $\Lambda^3 = (L H_0)^2 M_{\rm Pl}$, and the critical density defined as $\rho_{\rm crit} = 3M^2_{\rm Pl}H^2_0$ with the Hubble constant $H_0$.
Thus, in the core region, $A(r)$ saturates to a constant that is independent of the radius.
In the strongly screened regime, $A_{\rm core}\gg1$,
the deviation of the acceleration $\Delta_{\rm acc}$ in the core region is given by
\begin{align}
\Delta_{\rm acc}
\simeq
\begin{cases}
\displaystyle
\frac{2\xi}{3A_{\rm core}}
\sqrt{\frac{\xi}{3\beta}},
& (\beta>0),
\\[1ex]
\displaystyle
\frac{\xi^{4/3}}
{\bigl(\nu+2\alpha^{2}\bigr)^{1/3}}
A_{\rm core}^{-2/3},
& (\beta=0).
\end{cases}
\label{eq:Delta_core}
\end{align}
The magnitude of the residual fifth force is controlled by $A_{\rm core}$. Eq.~\eqref{eq:Delta_core} shows that, in the strongly screened regime $A_{\rm core}\gg1$, the residual is suppressed as $\Delta_{\rm acc} \propto A_{\rm core}^{-1}$ for $\beta>0$, or as $\Delta_{\rm acc} \propto A_{\rm core}^{-2/3}$ for $\beta=0$. Although $\Delta_{\rm acc}$ approaches a radius-independent value inside the core, its amplitude therefore becomes small when $A_{\rm core}$ is sufficiently large.

The value of $A_{\rm core}$ also determines whether a Vainshtein radius exists. Assuming that $A(r)$ does not exceed its central value $A_{\rm core}$, if $A_{\rm core}<1$, the condition $A(r_{\rm V})=1$ has no solution and no Vainshtein radius exists. In the limiting case $A_{\rm core}\ll1$, the solution remains close to the outer linear branch throughout the halo.

Conversely, if $A_{\rm core}>1$, continuity of $A(r)$ and its decay toward zero at large radii imply the existence of at least one Vainshtein radius. The existence of $r_{\rm V}$, however, does not guarantee that the fifth-force contribution vanishes toward the center, because $A(r)$ saturates at the finite value $A_{\rm core}$. When $A_{\rm core}\sim\mathcal{O}(1)$, including values moderately larger than unity, the central residual must be determined from the full algebraic equation, Eq.~\eqref{eq:varphi-eq}. A Vainshtein radius may therefore exist while an appreciable modified-gravity contribution remains in the central core.

To identify these regimes for representative cored halo profiles, let us introduce the core concentration parameter $c_s$ as
\begin{align}
c_s := \frac{R_{200}}{r_{\rm core}},
\end{align}
where $R_{200}$ denotes the radius below which the mean density becomes 200 times the critical density. Note that this definition of $c_s$ is useful to trace the mean density within the virial radius of a halo~(see, e.g., Ref.~\citep{Ogiya:2013qnf}).
We then assume that the corresponding mass $M_{200}=800\pi\rho_{\rm crit}R_{200}^3/3$ can be rewritten as
\begin{align}
M_{200}
=
4\pi \rho_{\rm core} r_{\rm core}^3 f_m(c_s),
\end{align}
Here $f_m(c_s)$ is a form factor that depends on the density profile and is a function of $c_s$ alone.
This yields
\begin{align}
\frac{\rho_{\rm core}}{\rho_{\rm crit}}
=
\frac{200}{3}
\frac{c_s^3}{f_m(c_s)}.
\label{eq:rho_core_cs}
\end{align}
Its definition and explicit expressions for representative halo models are summarized in App.~\ref{app:halo_profiles}.
Substituting Eq.~\eqref{eq:rho_core_cs} into Eq.~\eqref{eq:A_core_constant}, we obtain
\begin{align}
A_{\rm core}(L,c_s)
=
\frac{100}{3L^2}
\frac{c_s^3}{f_m(c_s)}.
\label{eq:A_core_L_cs}
\end{align}
Thus, the screening condition $A_{\rm core}\gg 1$ can be characterized on the $L$-$c_s$ plane. 

\begin{figure}[t]
\centering
\begin{minipage}{0.80\linewidth}
\centering
\includegraphics[width=\linewidth]{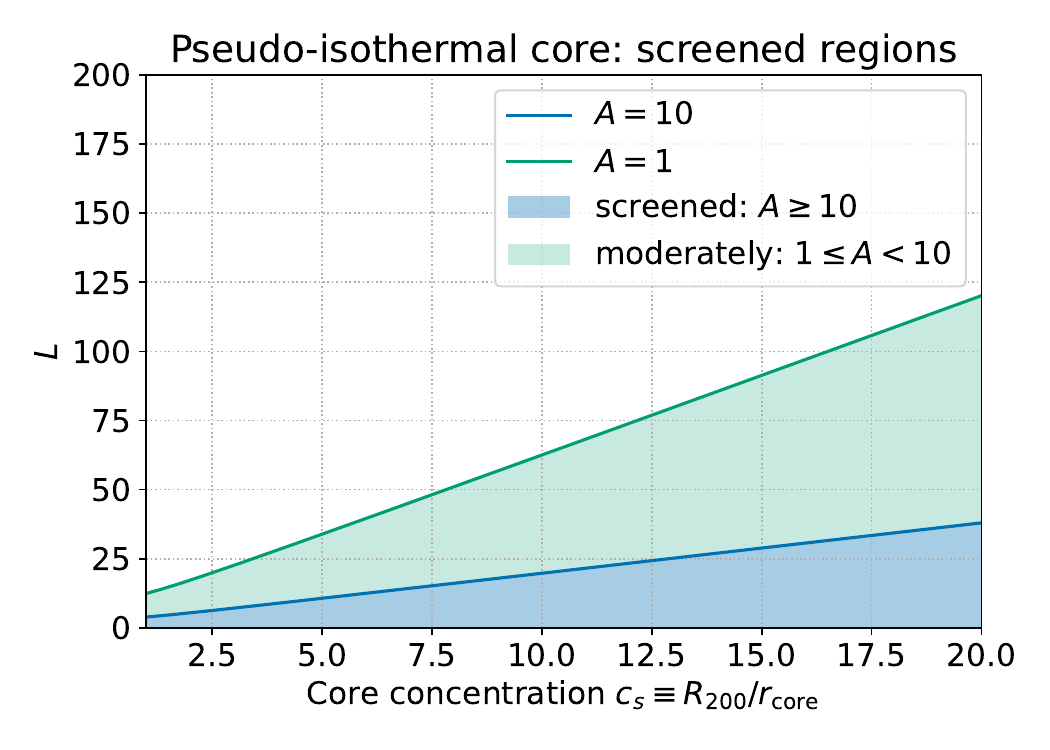}
\end{minipage}\hfill
\begin{minipage}{0.80\linewidth}
\centering
\includegraphics[width=\linewidth]{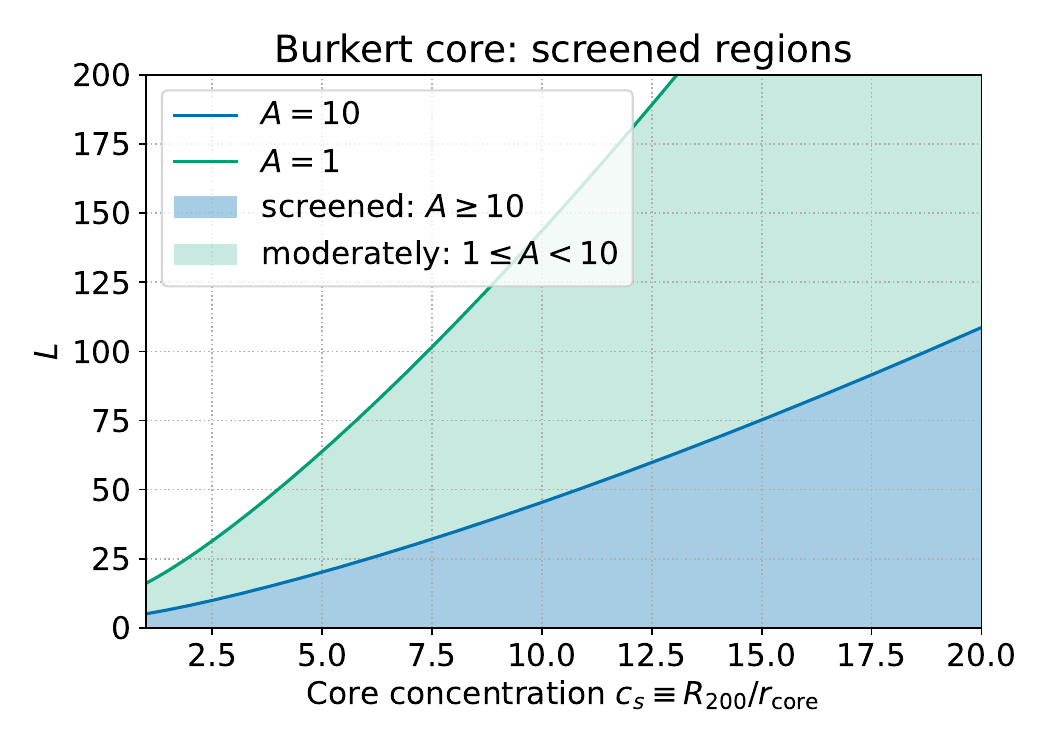}
\end{minipage}
\caption{
Examples of screening regimes in cored halos.
Top: pseudo-isothermal core. Bottom: Burkert core. In both panels, $A_{\rm core}$ is evaluated in the $L$-$c_s$ plane,
which is divided into the strongly screened
($A_{\rm core}\gg 1$), transitional
($A_{\rm core}\sim \mathcal{O}(1)$), and unscreened
($A_{\rm core}\ll 1$) regimes.
}
\label{fig:core_screen_regions}
\end{figure}

\begin{figure}[t]
\centering
\includegraphics[width=0.80\linewidth]{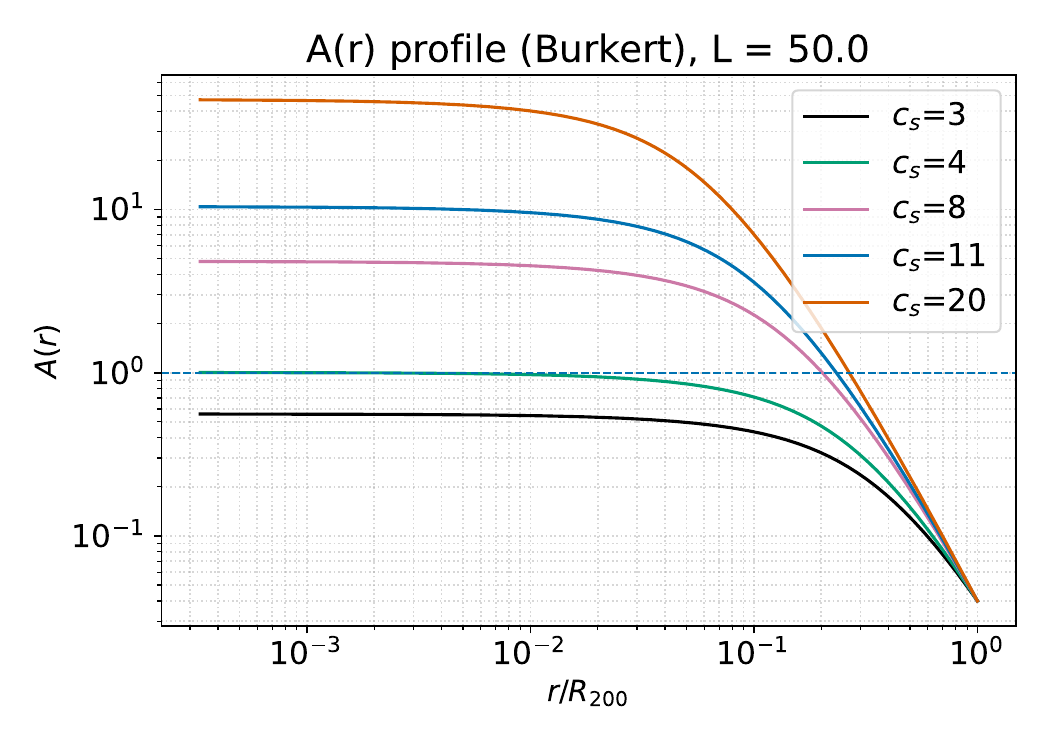}
\caption{
Example of $A(r)$ for a Burkert profile with ($L=50$).
The function $A(r)$ can saturate near the core radius, and the
saturation value $A_{\rm core}$ varies significantly with $c_s$.
}
\label{fig:A_profile}
\end{figure}

We aim to show how the domains in the $L$-$c_s$ plane are divided by the regions that characterize how effective the Vainshtein screening is, when a halo profile with a core structure is given. 
As illustrative examples, let us consider the pseudo-isothermal core profile~\cite{Jimenez:2002vy} and the Burkert profile~\cite{Burkert1995} (see more details of halo profiles in App.~\ref{app:halo_profiles}), both of which exhibit a core region at $r < r_{\rm core}$. Figure~\ref{fig:core_screen_regions} shows the classification of the parameter space according to $A_{\rm core}(L,c_s)$. We find that the region in which the Vainshtein screening is efficient extends over a wide range of $L$ as $c_s$ becomes larger. Otherwise, the screened region shrinks. Since a smaller $c_s$ means a larger core, our result implies that the Vainshtein screening is not necessarily effective for halos with extended core profiles, even for the spherically symmetric setup. 
Figure~\ref{fig:A_profile} illustrates the radial dependence of $A(r)$ for a Burkert profile with various values of $c_s$.
It clearly demonstrates that $A(r)$ in the core region becomes greater than unity for $c_s$ greater than four. 

In general, the concentration parameter $c_s$ characterizes the scale for generic density profiles of a halo below which they become less cuspy compared to the outside of the core radius.
As can be seen in Eq.~\eqref{eq:A_to_gamma},
Less-cuspy profiles with smaller $\gamma$ imply that the screening regions are shrinking to smaller scales. 
Then, we would like to note that the slope of the inner density profile is significant in determining whether the Vainshtein screening is effective or not. 
On the other hand, since the density profile of a halo is, in principle, determined by the gravitational interaction, it is necessary to solve for both the gravitational interaction and the density profile in a self-consistent manner.

In the following section, we shall employ
$N$-body simulations that solve the continuity equation and the Euler equation, given the gravitational field dynamics, to clarify whether the feature of the weakened screening survives even within the Vainshtein radius or not, numerically.

\section{$N$-body simulations for spherical haloes in Horndeski theory}
\label{sec:nbody_sim_spherical_haloes_Horndeski}

In this section, we describe a numerical study that clarifies how effective the Vainshtein screening is for halos, employing $N$-body simulations that trace the whole dynamical evolution.  For quantitatively assessing the Vainshtein screening, we adopt the following minimal setup of simulations:
\begin{itemize}
\item Consider an isolated, spherically symmetric halo.
\item The fifth force is treated as a central force determined by the enclosed mass, while the two-body interaction is Newtonian gravity.
\item The quasi-static spherically symmetric scalar-field solution given by Eq.~\eqref{eq:varphi-eq} is solved at each time step.
\end{itemize}
In particular, our goal is to verify, through self-consistent dynamical evolution of halo structure, the criterion discussed in the previous section: namely, that in cored halos, saturation of the enclosed mean density may weaken screening in the central region. To this end, the setup we use is justified.

\subsection{Code and coordinate system}
\label{ssec:code_and_coorinate_sysmte}

We use Gadget-2 \cite{Springel2005} to compute gravitational interactions and implement the additional acceleration from MG
(fifth force) as a user-defined external force.
We adopt physical coordinates (no cosmological expansion, $a=1$), do not use periodic boundary conditions, i.e., an isolated system, and focus on halo relaxation, stability, and structural changes induced by MG.

The internal unit system used in Gadget-2 is shown in Table~\ref{tab:gadget_units}. All simulation times are given in these internal units. The numerical values correspond to those specified in the parameter file for reproducibility.
\begin{table}[t]
  \centering
  \caption{Internal unit system adopted in Gadget-2 (as specified in the parameter file).}
  \label{tab:gadget_units}
  \begin{tabular}{lc}
    \hline
    Quantity & Value \\
    \hline
    Number of particles & $10^7$ \\
    Representative scale & $R_{200} \sim 2$Mpc \\
    Softening length $\epsilon$ & 1 {\rm kpc} \\
    \hline
    UnitLength\_in\_cm & $6.171355162982734\times10^{24}\,{\rm cm}$ \\
    UnitMass\_in\_g    & $3.674721708562721\times10^{48}\,{\rm g}$ \\
    UnitVelocity\_in\_cm\_per\_s & $1.9931958573429883\times10^{8}\,{\rm cm\,s^{-1}}$ \\
    \hline
    (Derived) UnitLength & $\simeq 2\,{\rm Mpc}$ \\
    (Derived) UnitMass   & $\simeq 1.85\times10^{15}\,M_\odot$ \\
    (Derived) UnitVelocity & $\simeq 1993\,{\rm km\,s^{-1}}$ \\
    (Derived) UnitTime $(=L/V)$ & $\simeq 0.98\,{\rm Gyr}$ \\
    \hline
  \end{tabular}
\end{table}
The gravitational softening length is
\texttt{SofteningHalo} $= 5\times 10^{-4}$ in internal length units, corresponding to $\epsilon \simeq 1\,{\rm kpc}$ in physical units. This value was chosen following empirical guidance for stable tracking of central halo structure in isolated simulations (see, e.g., \cite{Ogiya:2017hbr}).
\begin{table}[t]
  \centering
  \caption{Key Gadget-2 run parameters used in this work (excerpt).}
  \label{tab:gadget_runparams}
  \begin{tabular}{lc}
    \hline
    Parameter & Value \\
    \hline
    \texttt{ComovingIntegrationOn} & 0 (physical coordinates) \\
    \texttt{PeriodicBoundariesOn} & 0 (isolated) \\
    \texttt{TimeBegin} -- \texttt{TimeMax} & 0.0 -- 13.0 \\
    \texttt{ErrTolIntAccuracy} & 0.01 \\
    \texttt{CourantFac} & 0.15 \\
    \texttt{MaxSizeTimestep} & 0.001 \\
    \texttt{ErrTolTheta} & 0.3 \\
    \texttt{TypeOfOpeningCriterion} & 1 \\
    \texttt{ErrTolForceAcc} & 0.005 \\
    \texttt{SofteningHalo} & $5\times 10^{-4}$ (internal length) \\
    \texttt{SofteningHaloMaxPhys} & $5\times 10^{-4}$ (same) \\
    \hline
  \end{tabular}
\end{table}
Initial conditions are constructed as spherically symmetric halos in equilibrium under Newtonian gravity, using MAGI \cite{Miki2018} to generate particle and velocity distributions. MAGI constructs $N$-body initial conditions in phase space for a given density profile and assumed velocity model. 
Control runs with pure Newtonian gravity are also performed to confirm that the initial conditions remain stationary as isolated systems.
We adopt the Navarro-Frenk-White (NFW) profile~\cite{Navarro1997} and the Burkert core profiles~\cite{Burkert1995} for comparison. In App.~\ref{app:halo_profiles}, we summarize the halo profile used in this paper. The NFW profile is used as a reference for a representative profile of the dark matter halo in Newtonian gravity.
In the subsequent analysis, we vary the core concentration parameter $c_s \equiv R_{200}/ r_\star$ to systematically investigate the screening-efficiency reduction due to cores  (see Sec.~\ref{ssec:core_profile_analysis}). Here we take $r_\star =r_{\rm core}$ for Burkert and $r_\star =r_s$ for NFW profiles as the characteristic scale.

Representative parameters are summarized in Tables ~\ref{tab:mg_params} and \ref{tab:halo_models}.
Note that it has been known for the Horndeski theory that $G_5(\phi,X)$ is tightly constrained by the bound between the speeds of the electromagnetic waves and the gravitational waves measured by the simultaneous events of GW170817 and GRB170817A, leading $\alpha_T \simeq 0 $, where $\alpha_T$ denotes the dimensionless deviation of the speed of gravitational waves 
from the speed of light~\citep{2017ApJ...848L..13A, 2017PhRvL.119y1302C,2017PhRvL.119y1301B,2017PhRvL.119y1303S, 2018PhRvD..97j4038A}.
This fact
restricts ourselves to set
$\beta \simeq 0$, as shown in App.~\ref{app:Horndeski_halo_model_detailed}. Then one may wonder how significant it is to consider the models with $\beta \neq 0$. We answer 
this as follows. The aim of this paper is to clarify how the halo profile physically imprints any observable effects, extending the theoretical analysis in \cite{Narikawa2013} for a more realistic halo profile with a consistent treatment of the dynamical evolution of a spherical halo formation. In addition, any complementary constraints on $G_5(\phi, X)$ are significant, as the only multi-messenger event of GW170817 and GRB170817A currently provides constraints. Hence, we claim that our analysis is able to be used for complementary constraints on $G_5(\phi, X)$ by future observations of galaxy clusters.

\subsection{Numerical Implementation}
\label{ssec:numerical_implementation}

We now outline the algorithm used to incorporate the central-force fifth-force calculation into Gadget-2. We focus on definitions of physical quantities and computational steps rather than MPI or parallel implementation details.
At each time step, we define the halo center $\bm r_{\rm c}$ and compute particle radii $r_i = |\bm x_i - \bm r_{\rm c}|$. Sorting particles by $r_i$ and taking cumulative sums yields $M(<r_i)$.
We then evaluate the dimensionless source
\begin{align}
A(r)
=
\frac{1}{M_{\rm Pl}\Lambda^3}
\frac{M(<r)}{8\pi r^3}.
\label{eq:A_nbody}
\end{align}
The same enclosed mass is also used for evaluating the spherical Newtonian acceleration as a diagnostic, enabling direct comparison between Newtonian and fifth-force contributions.

Since $P(x; A)=0$ (Eq.~\eqref{eq:varphi-eq}) may have multiple real roots for a given $A$ (see Sec.~\ref{sec:vainshtein_in_horndeski_spherical_halo}), we must numerically follow the branch that smoothly connects the outer decaying solution to the inner Vainshtein solution.

Let us recall the case (i) $\beta >0$. Since $P(0,A)=\xi A>0$ and $P(x_-,A)<0$, there is at least one root in $(x_-,0)$. In the chosen parameter region, this root is effectively unique. We therefore fix the search interval to $x_{\min}=x_-\,,
x_{\max}=0$ and search for the root within this interval. The algorithm proceeds as follows:
\begin{itemize}
\item First attempt rapid convergence using the Newton method.
\item Use the solution from the previous time step $x^{(n-1)}(r)$ as the initial guess to maintain branch continuity.
\item If the Newton method fails to converge or exits the interval $[x_{\min},x_{\max}]$, switch to the bisection method, which guarantees convergence under the sign-change condition.
\end{itemize}
This Newton–bisection hybrid ensures both robustness and computational efficiency.
For $\beta=0$, Eq.~\eqref{eq:varphi-eq} reduces to a cubic function of $x$. We enumerate real roots and select the one continuously connected to the outer solution ($x<0$). In the parameter region considered, this root is unique.

\subsection{Fifth-force acceleration as central force}
\label{ssec:fifth_force_as_central_force}

We numerically treat the fifth-force acceleration as a central force, discretizing $\Phi'$ given in Eq.~\eqref{eq:Delta_acc_def} as
\begin{align}
\bm a_i^{(5)}
=
-\frac{\Lambda^3}{M_{\rm Pl}}
\left(
-\xi x(r_i)
+\beta x(r_i)^3
\right)
(\bm x_i - \bm r_{\rm c}).
\label{eq:a5_vec}
\end{align}
Here, the quantity in parentheses corresponds to the radial component $F_\phi/r$. The factor $\Lambda^3/M_{\rm Pl}$ is converted to Gadget-2 internal units before implementation.

To ensure numerical stability near the center, we set
\begin{align}
\bm a^{(5)} = 0
\quad \text{for} \quad r < \epsilon,
\end{align}
where $\epsilon$ is the softening length ($\simeq 1\,{\rm kpc}$).
We fix $\bm r_{\rm c}$ at the origin and monitor the halo center-of-mass displacement to ensure that the central-force approximation remains valid. %
At each time step, the fifth-force calculation proceeds as follows:
\begin{enumerate}
\item Determine halo center $\bm r_{\rm c}$.
\item Compute $r_i = |\bm x_i - \bm r_{\rm c}|$.
\item Sort particles by $r_i$ and compute $M(<r_i)$.
\item Evaluate $A(r_i)$ and solve $P(x;A)=0$ for $x_i$.
\item Compute $\bm a_i^{(5)}$ from Eq.~\eqref{eq:a5_vec} and add to the particle acceleration.
\end{enumerate}
\begin{table}[t]
  \centering
  \caption{Modified gravity parameters used in the simulations. }
  \label{tab:mg_params}
  \begin{tabular}{cc}
    \hline
    Parameter & Value \\
    \hline
    $\xi$          & $1$ \\
    $\alpha$       & $0.1$ \\
    $\beta$        & $0$\,, $0.1$ \\
    others         & $0$ \\
    $\Lambda^{3}=(LH_0)^2 M_{\mathrm{Pl}}$  & $(50 H_{0})^{2} M_{\mathrm{Pl}}$ \\
    \hline
  \end{tabular}
\end{table}
\begin{table}[t]
  \centering
  \caption{Halo models adopted in the simulations.}
  \label{tab:halo_models}
  \begin{tabular}{lccc}
    \hline
    Profile & $M_{\rm tot}$ & $R_{200}$ & $c_s$ \\
    \hline
    NFW                 & $1.85\times10^{15}\,M_\odot$ & $2\,\mathrm{Mpc}$ & $10.0$ \\
    Burkert ($c_s=4$)   & $1.85\times10^{15}\,M_\odot$ & $2\,\mathrm{Mpc}$ & $4.0$ \\
    Burkert ($c_s=8$)   & $1.85\times10^{15}\,M_\odot$ & $2\,\mathrm{Mpc}$ & $8.0$ \\
    Burkert ($c_s=20$)  & $1.85\times10^{15}\,M_\odot$ & $2\,\mathrm{Mpc}$ & $20.0$ \\
    \hline
  \end{tabular}
\end{table}


%
%

\subsection{The behavior of $A(r)$, $x(r)$ and the gravitational acceleration in Horndeski theory}
\label{ssec:behavior_A_x}

\begin{figure*}[tbp]
\centering
\begin{subfigure}[t]{0.48\textwidth}
\centering
\includegraphics[width=0.8\linewidth]{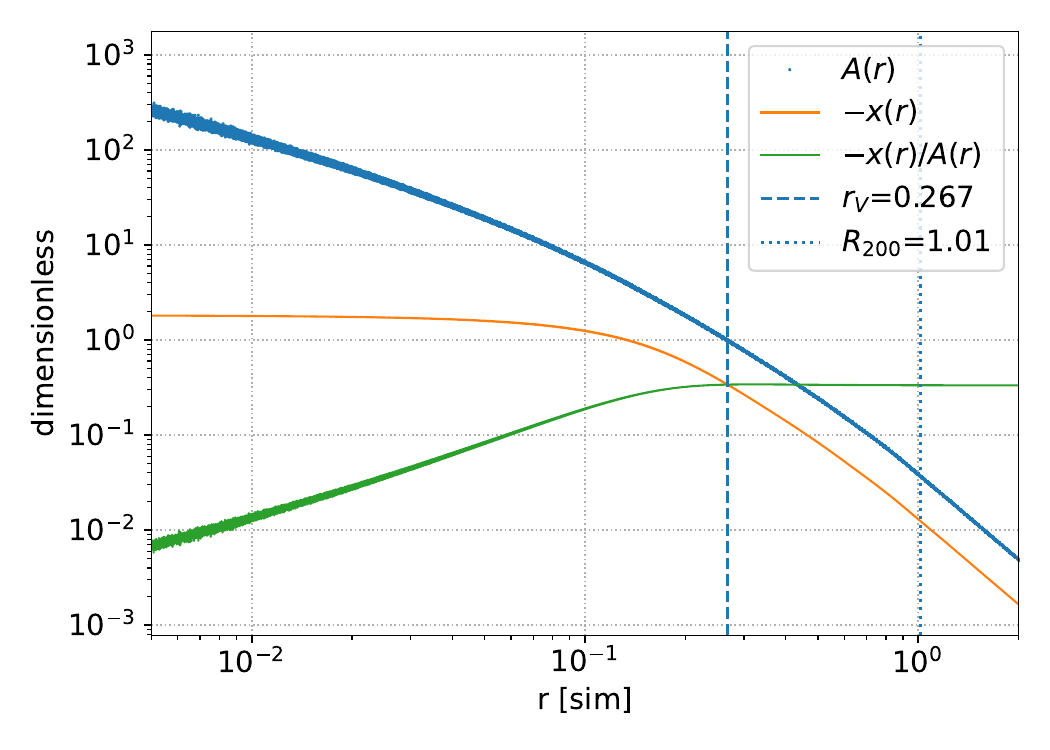}
\caption{NFW (representative case)}
\label{fig:Ax_nfw}
\end{subfigure}
\hfill
\begin{subfigure}[t]{0.48\textwidth}
\centering
\includegraphics[width=0.8\linewidth]{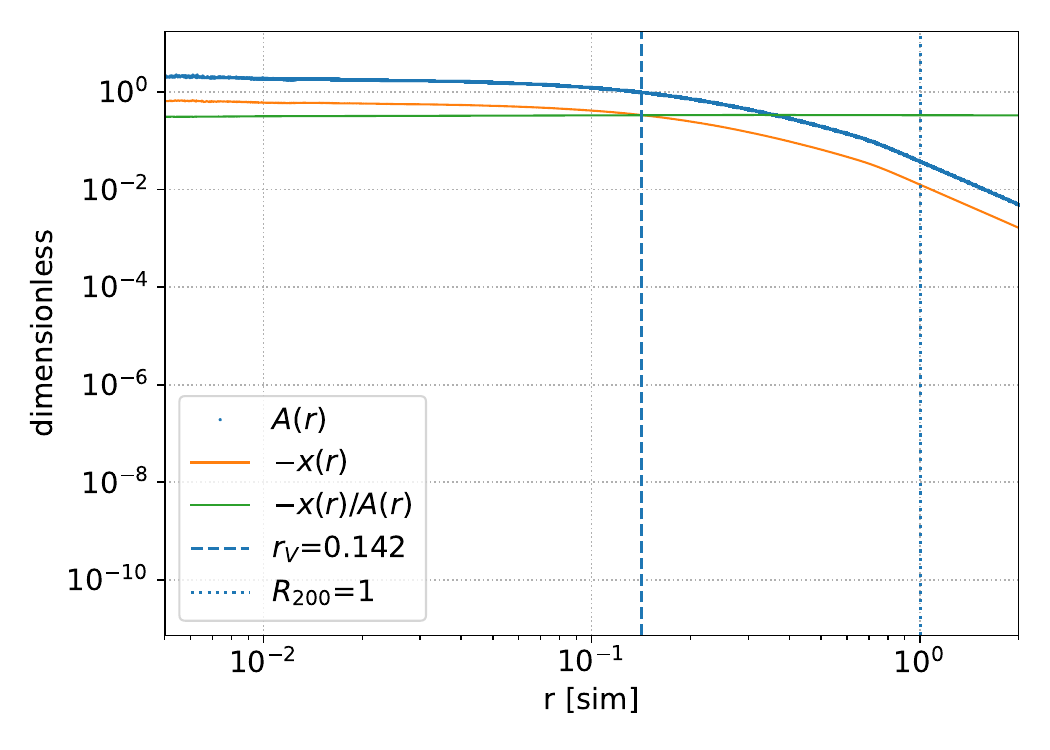}
\caption{Burkert: $c_s=4$}
\label{fig:Ax_cs4}
\end{subfigure}
\vspace{4mm}
\begin{subfigure}[t]{0.48\textwidth}
\centering
\includegraphics[width=0.8\linewidth]{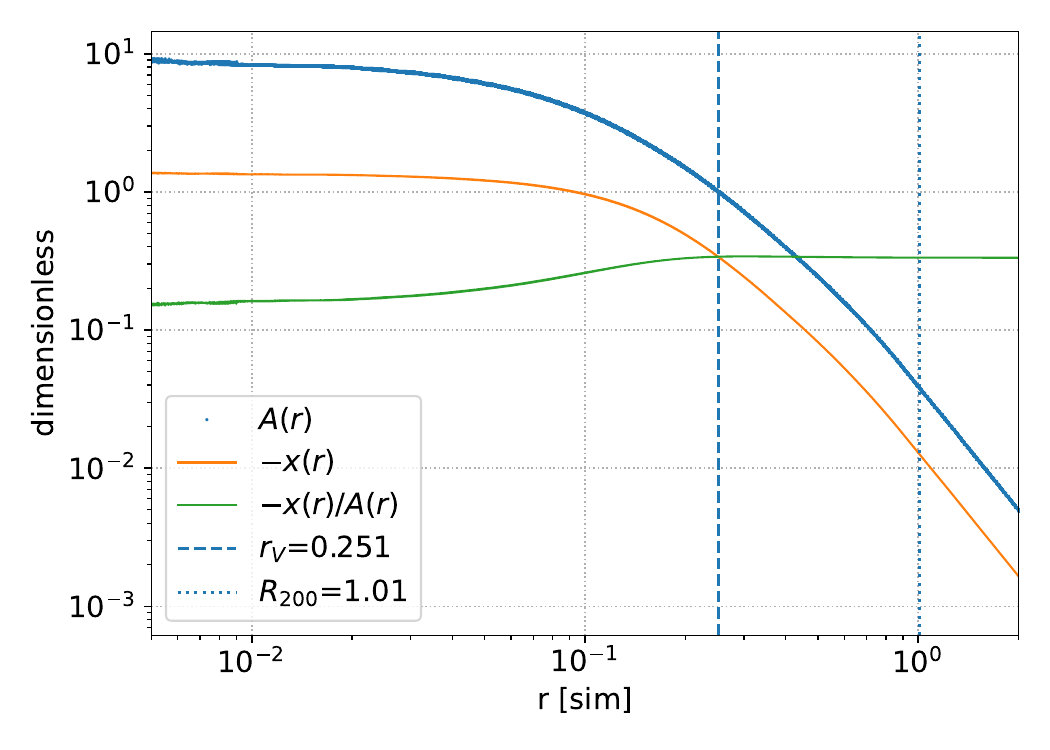}
\caption{Burkert: $c_s=8$}
\label{fig:Ax_cs8}
\end{subfigure}
\hfill
\begin{subfigure}[t]{0.48\textwidth}
\centering
\includegraphics[width=0.8\linewidth]{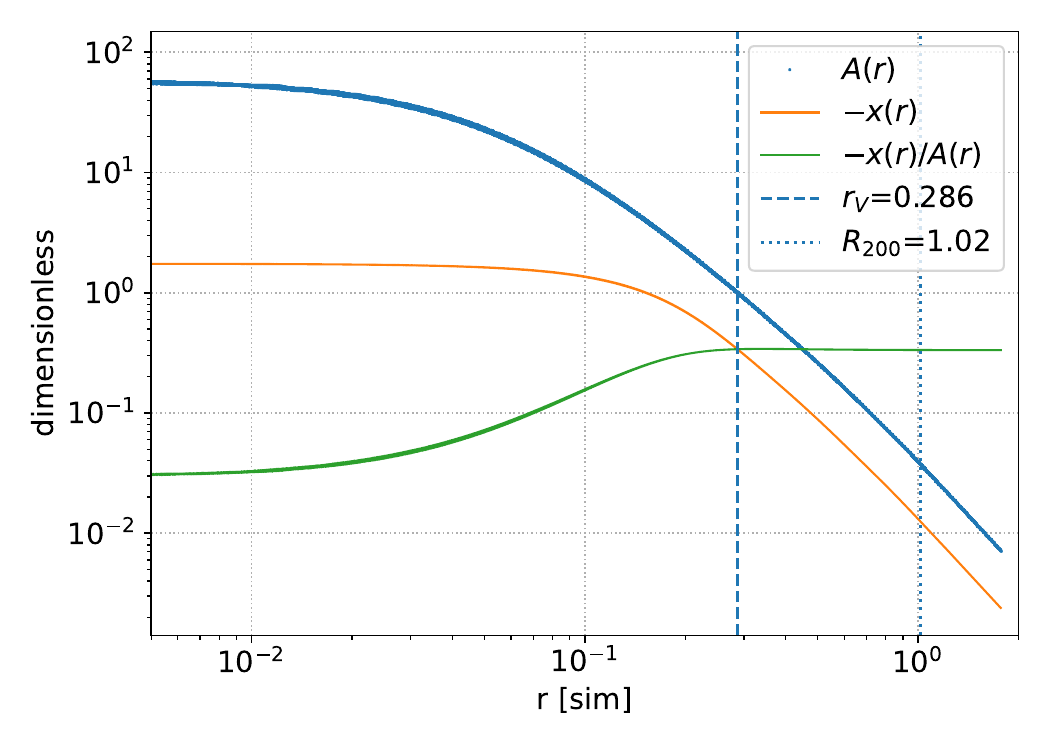}
\caption{Burkert: $c_s=20$}
\label{fig:Ax_cs20}
\end{subfigure}
\caption{
Profiles of the dimensionless source $A(r)$ and the scalar-field variable $x(r)$
for each model at snapshot $800$.
The blue points show $A(r)$, the orange line shows $-x(r)$, and the green line
shows $-x(r)/A(r)$. The dashed and dotted lines indicate $r_{\rm V}$ and $R_{200}$, respectively.
The unit ``[sim]" on the horizontal axis is $2~{\rm Mpc} = R_{200}$.
}
\label{fig:four_plots_subfig}
\end{figure*}

In Fig.~\ref{fig:four_plots_subfig}, we show the profiles of the function $A(r)$ and the scalar field $x(r)$ for each model of the density profile, comparing the NFW profile and the Burkert profiles with the three different core parameters: $c_s = 4, 8, 20$, respectively. We consider the halos that satisfies $R_{200} > r_{\rm V}$ with $A(r_{\rm V}) = 1$.
$A(r)$ is the most cuspy in the NFW profile, and it gets closer to the core profile when the core parameter $c_s$ becomes smaller. Correspondingly, the Vainshtein radius derived from the condition $A(r_{\rm V}) = 1$ becomes smaller as $c_s$ does. Note that the Vainshtein radii of the NFW profile and the Burkert profile with $c_s = 20$ are almost identical even though the inner core structures of $A(r)$ quantitatively differ. $x(r)$ increases and saturates at a constant value when it crosses the Vainshtein radii. The values to which $x(r)$ saturates are similar in order among all the density profiles. Let us recall how $\Delta_{\rm acc}$ scales with $A(r)$ and $x(r)$ in Sec.~\ref{ssec:scaling_of_Delta_acc}. In the inner regions where $A(r)\gg 1$ is satisfied, $\Delta_{\rm acc} \propto 1/A(r)$ and thus the Vainshtein screening suppresses the fifth force. As we describe in Sec.~\ref{fig:core_screen_regions}, the smaller $c_s$ becomes, the less efficient the Vainshtein screening becomes. The numerical implementation here quantitatively shows this behavior. The other quantity $-x(r)/A(r)$ traces the strength of $\Delta_{\rm acc}$ in the outer region $A(r) \ll 1$, extrapolated from the outer to the inner radii. We find that the NFW profile and the Burkert profile of $c_s = 20$ exhibit different transitions of $-x(r)/A(r)$, which may imply that the dynamical infall of the matter traces this feature before and after crossing the Vainshtein radius.
This motivates us to make use of the proxies of acceleration, namely the dynamical mass proxies. In the next section, we introduce the dynamical mass proxies and their traceability of the history of acceleration.
%

\section{Multiple mass proxies}
\label{sec:mass_proxies}
\vspace{-0.1cm}
In this section, we shall introduce multiple mass proxies for dark matter halos, clarifying potential signatures of MG that can be obtained from observations as systematic biases of the mass proxies. We employ multiple mass-estimation methods that are commonly used in galaxy-cluster observations, i.e.,  hydrostatic equilibrium (HSE)~\cite{1976A&A....49..137C, KomatsuSeljak2001}, Jeans analysis~\citep{1915MNRAS..76...70J}, and weak gravitational lensing~\cite{Tyson1990, KaiserSquires1993}. We numerically compute the mass proxies given the three methods for an isolated spherically symmetric halo via $N$-body simulation outputs. In post-processing the outputs of the simulation, we quantify the degree to which inferred masses are systematically biased, given that all the mass proxies are consistent with Newtonian gravity.
In what follows, we describe how the three mass proxies can trace the effect of MG. 

\subsection{True mass}
\label{ssec:true_mass}

Let us define the ``true enclosed mass'' directly estimated from the $N$-body dark matter particle distribution $\rho_{\rm dm}$ as
\begin{align}
M_{\rm true}(<r)
:=
\int_0^r 4\pi r'^2\,\rho_{\rm dm}
(r')\,{\rm d}r'.
\label{eq:Mtrue_def_mass}
\end{align}
Note that the definition of true mass in Eq.~\eqref{eq:Mtrue_def_mass} is reasonable, as far as we work in the Jordan frame in which the matter is minimally coupled to gravity.

\subsection{Dynamical mass proxies}
\label{ssec:mass_definitions}

Based on Newton's equation of motion, we can define the ``dynamical mass" as
\begin{align}
M_{\rm dyn}(<r):=\frac{r^2}{G_{\rm N}}\,g(r),
\label{eq:Mdynamical_def_mass}
\end{align}
from the gravitational acceleration, $g(r)$.
Of course, when the gravitational acceleration is driven only by Newtonian force, the identity $M_{\rm true} = M_{\rm dyn}$ follows at any radius $r$. 
Provided that MG puts an extra acceleration as the fifth force, on the contrary, the acceleration felt by particles is
\begin{align}
g(r)=g_{\rm N,true}(r)+g_{5}(r),
\qquad
g_{\rm N,true}(r)=\frac{G_{\rm N}M_{\rm true}(<r)}{r^2}.
\label{eq:g_total_mass}
\end{align}
Thus, when an observer assumes Newtonian gravity to reconstruct $M_{\rm dyn}$, it does not generally coincide with $M_{\rm true}$. We define the deviation as
\begin{align}
\Delta_{\rm acc}(r)
:=
\frac{g(r)}{g_{\rm N,true}(r)}-1
=
\frac{M_{\rm dyn}(<r)}{M_{\rm true}(<r)}-1\,.
\label{eq:Delta_acc_def_mass}
\end{align}
Note that Eq.~\eqref{eq:Delta_acc_def_mass} is equivalent to the parameter in Eq.~\eqref{eq:Delta_acc_def}, given that the fifth-force acceleration is determined by the Horndeski theory.

There are two representative mass proxies to measure $M_{{\rm dyn}}$, introducing the HSE mass proxy $M_{\rm HSE}$ and the Jeans mass proxy $M_{\rm Jeans}$ as follows.

\subsubsection{Hydrostatic equilibrium mass proxy}
\label{sssec:HSE_mass}

Let us assume that the baryon gas in a certain region in a halo reaches hydrostatic equilibrium and is spherically symmetric. Then the gas pressure and gravity are in balance~\cite{Terukina2014} as
\begin{align}
\frac{dP_{\rm gas}}{dr}
=
-\rho_{\rm gas}(r)\,g(r).
\label{eq:HSE_basic_mass}
\end{align}
Thus, assuming the equation of state for the baryon gas in the halo, it is possible to derive information about $g(r)$ from the temperature and density distribution of the gas using this equation.
For an ideal gas,
\begin{align}
P_{\rm gas}
=
\frac{k_{\rm B}T_{\rm gas}}{\mu m_p}\rho_{\rm gas}\,,
\label{eq:ideal_gas_mass}
\end{align}
with Boltzmann constant $k_{\rm B}$, proton mass $m_p$, and the mean molecular weight, $\mu$,
from Eq.~\eqref{eq:HSE_basic_mass}, the mass proxy can be estimated as
\begin{equation}
M_{\rm HSE}(<r)=
-\frac{r}{G_{\rm N}}
\frac{k_{\rm B}T_{\rm gas}(r)}{\mu m_p}
\left[
\frac{d\ln\rho_{\rm gas}}{d\ln r}
+\frac{d\ln T_{\rm gas}}{d\ln r}
\right]\,.
\label{eq:MHSE_standard_mass}
\end{equation}
In practice, the estimation of $M_{\rm HSE}(<r)$ is made by deriving information on temperature and density distributions from observations. However, in our analysis, we just construct a gas distribution based on the output from $N$-body simulations, such as the gravitational acceleration $g(r)$, based on Eq.~\eqref{eq:HSE_basic_mass}, and then it is difficult to obtain the information on temperature and density independently. Then,
for simplicity, we introduce the polytropic relation~\cite{KomatsuSeljak2001}:
\begin{align}
P_{\rm gas}(r)
=
K_{\rm poly}\rho_{\rm gas}^\Gamma(r),
\label{eq:polytrope_mass}
\end{align}
and we use a further simplified expression as
\begin{align}
M_{\rm HSE}(<r)=
-\frac{r}{G_{\rm N}}
K_{\rm poly}\Gamma \exp\left[(\Gamma-1) y \right]
\frac{dy}{d \ln r}\,,
\label{eq:MHSE_standard_mass}
\end{align}
with $y := \ln \rho_{\rm gas}$. We compute $y$ from the value of $g(r)$ in the simulation, by integrating Eq.~\eqref{eq:HSE_basic_mass} given Eq.~\eqref{eq:polytrope_mass}.
We assume $\Gamma = 1.2$ following Ref.~\cite{Solanes:2005qk} and $K_{\rm poly}$ is specified from the outer boundary of the halo. 

\subsubsection{Jeans mass proxy}
\label{sssec:Jeans_mass}

We use the spherical Jeans equation to infer mass from the radial velocity dispersion profiles, assuming a steady state with no mean streaming
motion~\cite{Falco2013}. After averaging the velocity dispersion tensor over the sphere within a radius $r$, we obtain the Jeans equation as
\begin{align}
\frac{d(\rho_{\rm dm}\sigma_r^2)}{dr}
+\frac{2\beta_{\rm halo}(r)}{r}\rho_{\rm dm}\sigma_r^2
=
-\rho_{\rm dm}(r)\,g(r).
\label{eq:Jeans_basic_mass}
\end{align}
Here, $\sigma^2_r$ is the radial velocity dispersion, and $\beta_{\rm halo}(r)=1-\sigma^2_t/(2\sigma^2_r)$ is the velocity anisotropy parameter, where the tangential velocity dispersion is defined by $\sigma^2_t =\sigma_\theta^2+\sigma_\phi^2$. Making use of Eq.~\eqref{eq:Mdynamical_def_mass}, we obtain 
\begin{align}
M_{\rm Jeans}(<r)
=
-\frac{r\sigma_r^2(r)}{G_{\rm N}}
\left[
\frac{d\ln (\rho_{\rm dm}\sigma^2_r)}{d\ln r}
+2\beta_{\rm halo}(r)
\right].
\label{eq:MJeans_standard_mass}
\end{align}
We measure $\sigma^2_r$ and $\sigma^2_t$ from a snapshot of the $N$-body simulation, and compute $M_{\rm Jeans}$ by using Eq.~\eqref{eq:MJeans_standard_mass}.

\subsection{Mass proxy from weak gravitational lensing}
\label{ssec:WL_mock}

As another mass proxy, we introduce a mass estimate from the weak gravitational lensing effect, which is based on geometric information (see, e.g., Refs.~\cite{Tyson1990, KaiserSquires1993, Brainerd1996, Fischer2000}).

Introducing the polar coordinate $(R,\phi)$ with the center of the cluster as the origin, we define the lensing convergence $\kappa$ and the tangential shear field $\gamma_t$ as 
\begin{align}
\kappa(R)=\frac{\Sigma(R)}{\Sigma_{\rm crit}}\,,\ 
\gamma_t(R)=\bar{\kappa}(<R)-\kappa(R),
\label{eq:gamma_t_def_mass}
\end{align}
where $\Sigma (R)$, $\Sigma_{\rm crit}$, $\bar{\kappa}(<R)$ denote the surface mass density field, critical surface mass density, and the convergence averaged within a circle of radius $R$.
Under the thin-lens approximation, the convergence can be approximated as
\begin{align}
\kappa(R)\propto \int dZ\,\nabla^2\Phi_{\rm Weyl}(r),
\qquad
\Phi_{\rm Weyl} = \frac{\Phi+\Psi}{2}\,,
\end{align}
where $Z$ denotes the line-of-sight distance and $r=\sqrt{R^2+Z^2}$. In the Horndeski theory, using Eqs.~\eqref{eq:Phi-eq} and~\eqref{eq:Psi-eq}, one finds that the Weyl potential satisfies the following equation~\cite{Narikawa2013},
\begin{align}
\nabla^2\Phi_{\rm Weyl}(r)
=
\frac{\Lambda^3}{M_{\rm Pl}}
\frac{\left[\left(\alpha x^2+2\beta x^3+2A\right)r^3\right]'}{2r^2}.
\label{eq:lensing_laplacian_horndeski_mass}
\end{align}

The term proportional to $A$ corresponds to the standard contributions from the matter distributions,
whereas the terms involving
$\alpha$ and $\beta$ represent the direct contributions from MG.
We therefore decompose the convergence as
\begin{align}
  \kappa(R)=\kappa_{\rm N}(R)+\kappa_{5}(R).
\end{align}
Eq.~\eqref{eq:lensing_laplacian_horndeski_mass} suggests that 
MG does not produce a substantial global effect on weak lensing when
$\alpha$ and $\beta$ are small or when the variation of $x'(r)$ is
localized or small.

In GR, the Weyl potential can be related to the enclosed mass as
\begin{align}
\nabla^2\Phi_{\rm Weyl}(r)
= G_{\rm N} \frac{\left[M(<r)\right]^\prime}{r^2} ~,  
\end{align}
and then, the mass proxy based on the weak lensing measurement, $M_{\rm WL}(<r)$, is evaluated from the convergence as
\begin{align}
\kappa(R)\propto \int dZ\,G_{\rm N} \frac{\left[M_{\rm WL}(<r)\right]^\prime}{r^2}.
\label{eq:WL_standard_mass}
\end{align}

While the above is a rough sketch of the estimation, when actually calculating estimates based on the $N$-body simulations, we need to construct the lensing quantities from the $N$-body particle distribution and estimate the azimuthally averaged shear $\hat{\gamma}_t(R)$ from a synthetic catalog including shape noise. In Appendix~\ref{app:lensing}, we summarize the procedure we adopted for estimating the weak lensing mass proxy.

\section{Results}
\label{sec:results}

In this section, we show the results for the mass proxies introduced in the previous section, and discuss how the signatures of MG can be seen for each mass proxy, depending on the density profiles of halos.
Note that we measure $M_{\rm true}$ from simulations by using Eq.~\eqref{eq:Mtrue_def_mass}.
%

%

\begin{figure*}[htbp]
\centering
\begin{subfigure}[t]{0.48\textwidth}
\centering
\includegraphics[width=0.8\linewidth]{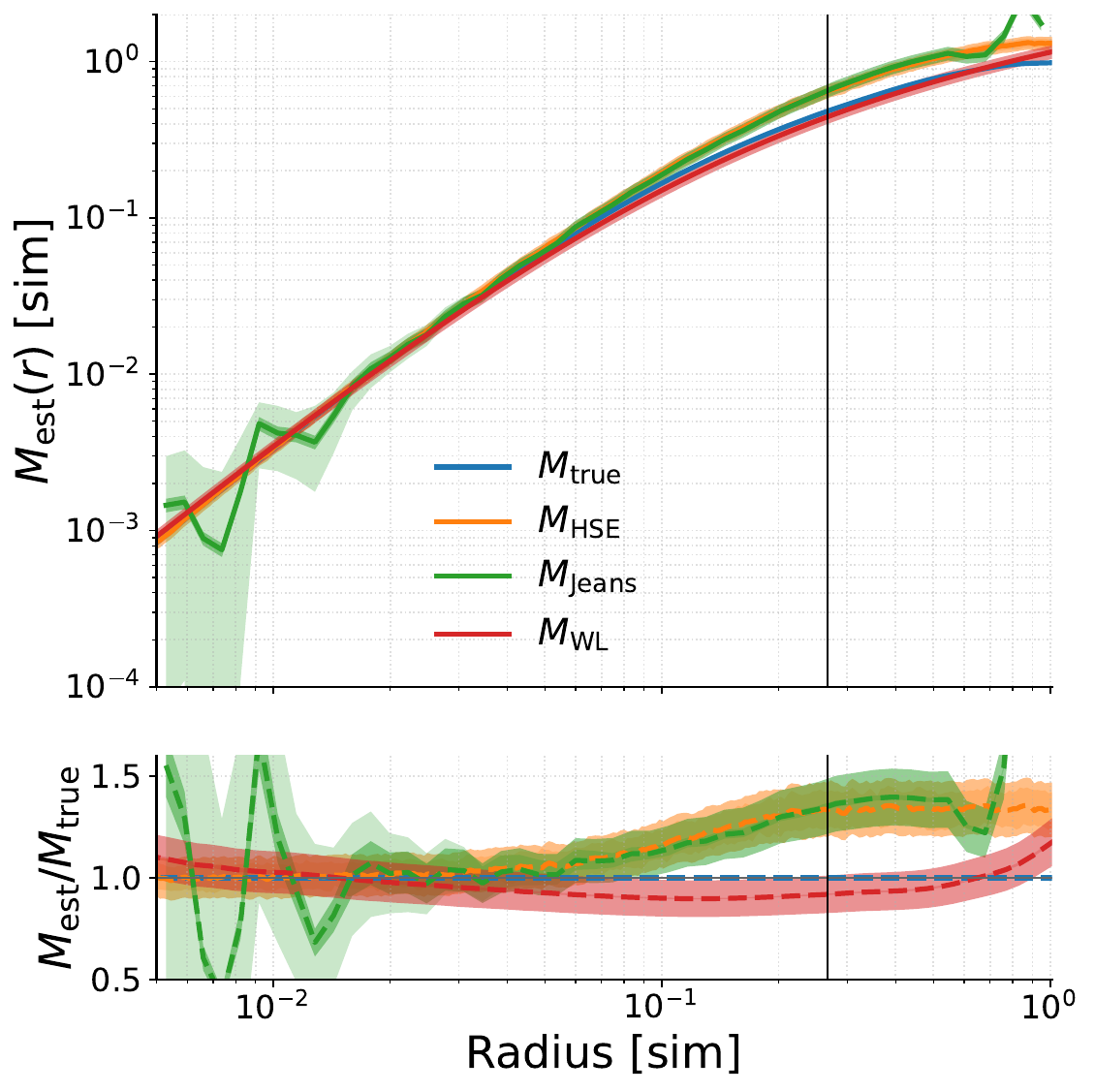}
\end{subfigure}
\hfill
\begin{subfigure}[t]{0.48\textwidth}
\centering
\includegraphics[width=0.8\linewidth]{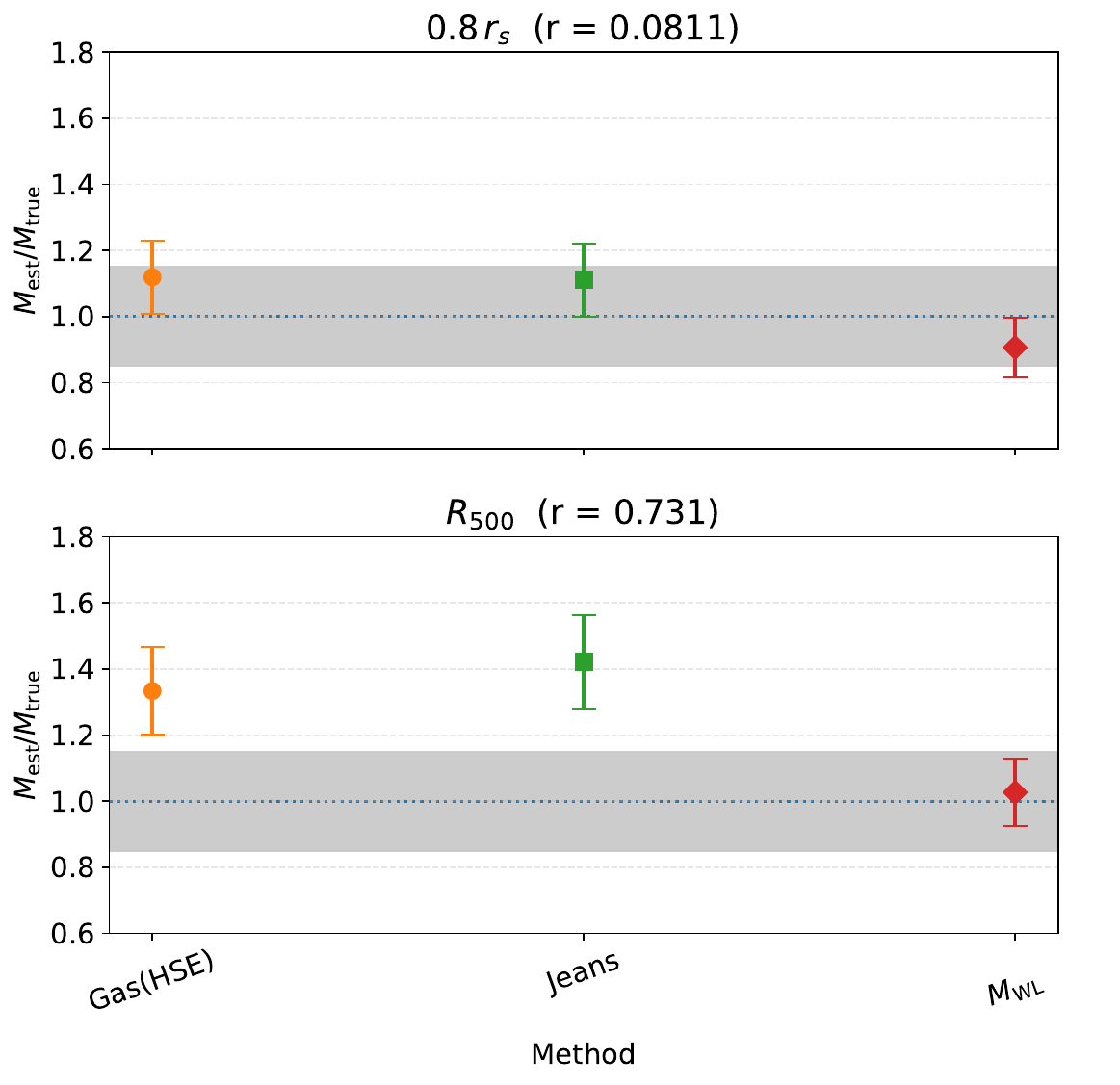}
\end{subfigure}
\caption{
Estimated mass proxies and true mass profiles in the NFW model at snapshot 800. The black solid vertical line denotes the Vainshtein radius (left panel).
Ratio of each mass proxy to the true mass at the representative radii with $r=0.8\,r_s$ and $R_{500}$ (right panel).
Each mass proxy is displayed with an artificial $10\%$ error, purely as a guide.
The unit ``[sim]" on the horizontal axis is $2~{\rm Mpc} = R_{200}$.}
\label{fig:mass_proxies_NFW}
\end{figure*}

Figure.~\ref{fig:mass_proxies_NFW} shows the mass estimates for the NFW halo. In the inner region, $r \lesssim r_{\rm V}$, $M_{\rm HSE}$ and $M_{\rm Jeans}$ are broadly consistent with $M_{\rm true}$. This behavior agrees with the analytical expectation that gravity approaches GR toward the halo center as the screening mechanism becomes effective. In particular, at the representative radius near $0.8r_s$, both $M_{\rm HSE}$ and $M_{\rm Jeans}$ deviate from $M_{\rm true}$ by only $\mathcal{O}(10\%)$, indicating that screening operates effectively in the vicinity of the halo center.

In the innermost few bins, however, the limited number of particles and the numerical instability associated with discrete differentiation can produce spike-like fluctuations, particularly in $M_{\rm Jeans}$. Since the primary objective of this study is to identify systematic biases arising from MG,
we hereafter focus mainly on the sufficiently well-resolved radial range, approximately $r \gtrsim 10^{-2}R_{200}$.

In the outer region,  $r \gtrsim r_{\rm V}$, both $M_{\rm HSE}$ and $M_{\rm Jeans}$ systematically overestimate $M_{\rm true}$. This overestimation is consistent with the enhancement of the gravitational acceleration in the outer solution,
\begin{align}
\Delta_{\rm acc}
\simeq
\frac{2\xi^2}{\eta+6\xi^2},
\end{align}
as given in Eq.~\eqref{eq:Delta_outer}. For the representative parameter choice adopted in this study, e.g., $\xi=1$ and $\eta=0$, one expects $\Delta_{\rm acc}\simeq 1/3$. If an observer assumes GR and reconstructs the dynamical mass according to $M_{\rm dyn}\propto g$, the resulting overestimation is
\begin{align}
\frac{M_{\rm HSE}}{M_{\rm true}}
\sim
\frac{M_{\rm Jeans}}{M_{\rm true}}
\sim
1+\Delta_{\rm acc}
\simeq
\frac{4}{3},
\end{align}
follows naturally. Indeed, the summary results around $R_{500}$ show that $M_{\rm HSE}$ and $M_{\rm Jeans}$ tend to be approximately $30\%$ higher than the true mass.

\begin{figure*}[htbp]
\centering
\begin{subfigure}[t]{0.48\textwidth}
\centering
\includegraphics[width=0.8\linewidth]{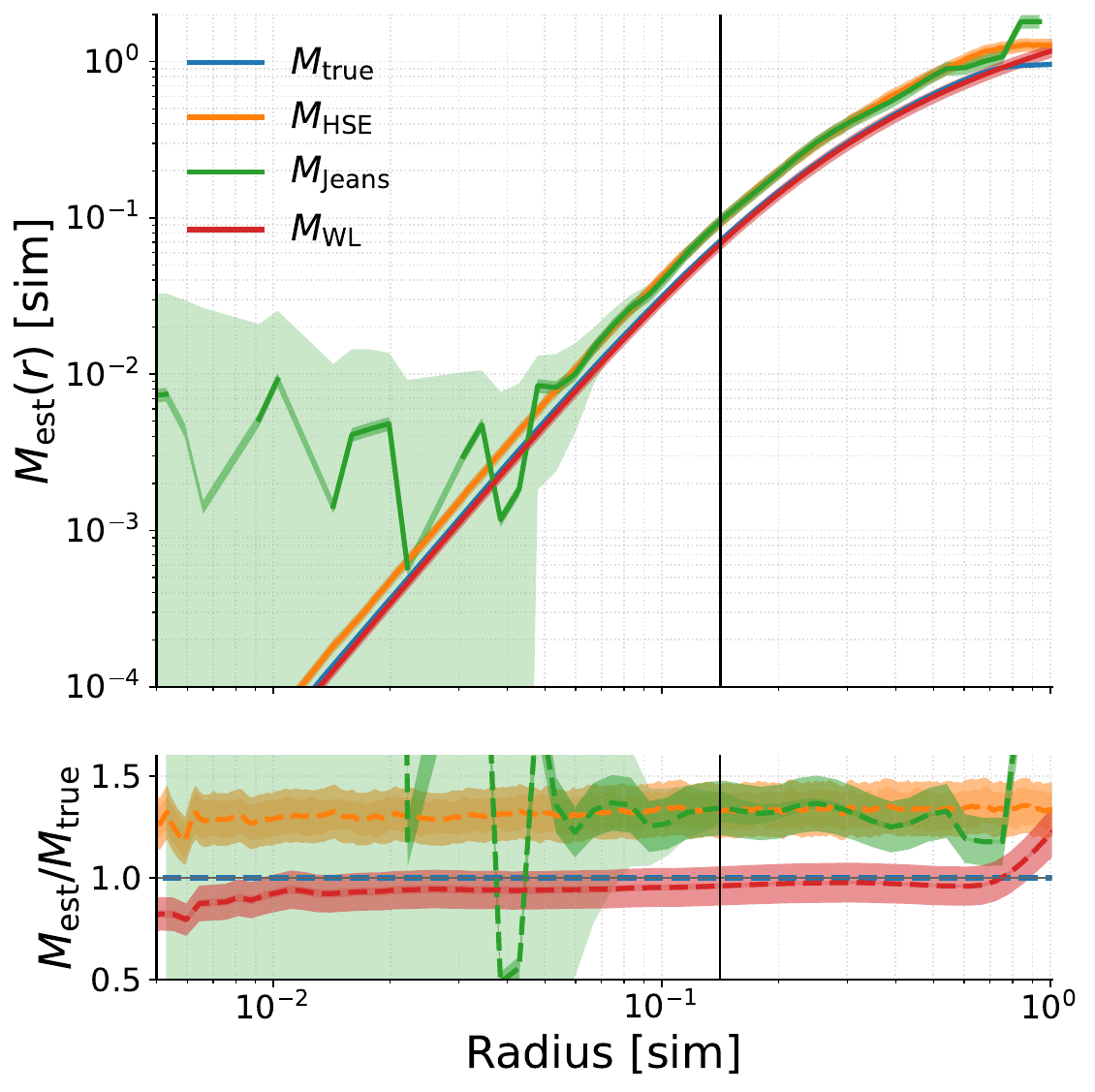}
\end{subfigure}
\hfill
\begin{subfigure}[t]{0.48\textwidth}
\centering
\includegraphics[width=0.8\linewidth]{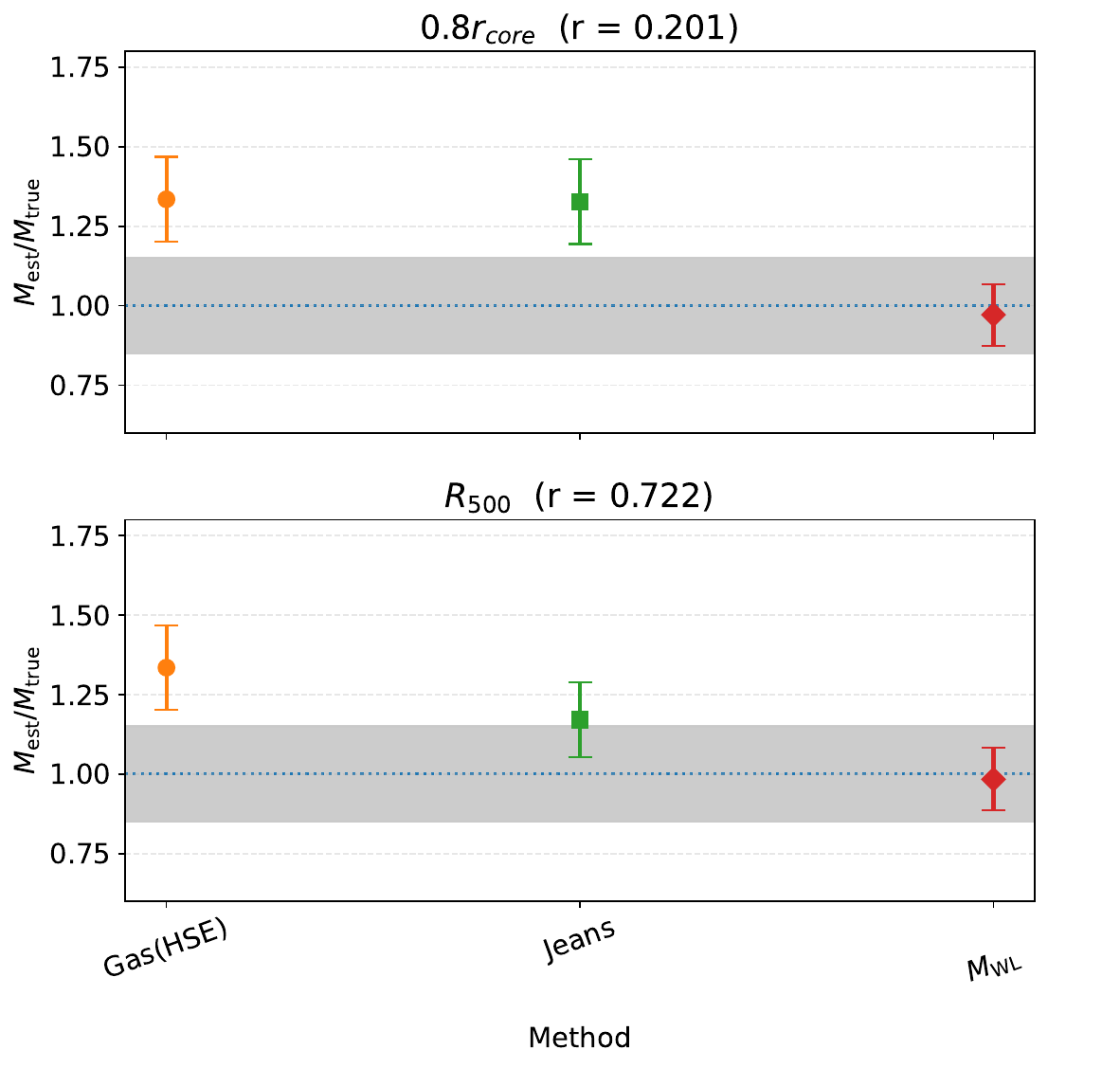}
\end{subfigure}
\caption{
Same plots in the Burkert cored model with $c_s = 4$ at snapshot 800. The black solid vertical line denotes the Vainshtein radius (left panel).
Ratio of each mass proxy to the true mass at the representative radii with $r=0.8\,r_{\rm core}$ and $R_{500}$ (right panel).
Each mass proxy is displayed with an artificial $10\%$ error, purely as a guide.
The unit ``[sim]" on the horizontal axis is $2~{\rm Mpc} = R_{200}$.
}
\label{fig:mass_proxies_burkert}
\end{figure*}

Figure ~\ref{fig:mass_proxies_burkert} shows the results for the Burkert core profile $c_s = 4$, with which the density profile exhibits the widest core region within the Vainshtein radius. In  the right panels, $0.8 r_{\rm core}$ and $R_{500}$ are adopted as representative radii. Recall that $A(r)$ does not become sufficiently large even toward the center, and the screening effect can saturate even in the inner region as we describe in Sec.~\ref{ssec:behavior_A_x}. In this case, $M_{\rm HSE}$ or $M_{\rm Jeans}$ consistently overestimate $M_{\rm true}$ from the central to the outer regions, demonstrating that a dynamical equilibrium state can be realized in which screening remains insufficient even within the halo. This result supports, through self-consistent $N$-body structure formation, one of the central conclusions of this study: MG
effects can persist toward the halo center because the presence of a core causes $A(r)$ to saturate.

In both Figs.~\ref{fig:mass_proxies_NFW} and \ref{fig:mass_proxies_burkert}, $M_{\rm WL}$, by contrast, reproduces $M_{\rm true}$ well overall. Depending on radius, deviations of approximately $10\%$ toward either lower or higher values are present, but this behavior is clearly distinct from the systematic overestimation of approximately $30\%$ found for $M_{\rm HSE}$ or $M_{\rm Jeans}$. The reason why $M_{\rm WL}$ remains relatively close to $M_{\rm true}$ can be understood from the different metric potentials probed by dynamical and lensing observables. While the dynamical mass proxies respond directly to $\Phi$, weak gravitational lensing probes the Weyl potential $\Phi_{\rm Weyl}$.
In the Weyl potential, the leading scalar-field contributions proportional to $\xi x$ cancel exactly. In particular, in the outer branch where $x\propto A$, one finds parametrically
\begin{align}
    \Delta_{\rm acc}=\mathcal{O}(1),
    \qquad
    \frac{\delta(\nabla^2\Phi_{\rm Weyl})}
         {\nabla^2\Phi_{\rm Weyl}^{\rm GR}}
    =
    \mathcal{O}(\alpha A)+\mathcal{O}(\beta A^2).
\end{align}
The fifth force can therefore produce an appreciable bias in $M_{\rm HSE}$ and $M_{\rm Jeans}$ while leaving a substantially smaller direct imprint on $M_{\rm WL}$. Moreover, the dominant matter contribution to the convergence is constructed from the actual particle distribution evolved under MG,
so that the corresponding structural change of the halo is shared by both $M_{\rm true}$ and $M_{\rm WL}$. These properties naturally explain why $M_{\rm WL}$ stays closer to $M_{\rm true}$ in Figs.~4 and 5.

%

\section{Conclusion and Discussion}\label{sec:conclusion}

The models of modified gravity that can explain the late-time cosmic acceleration must be examined through
cosmological observations of dark matter haloes, as well as the local tests of gravity. To pass any observational test of gravity, it is essential to verify whether the screening mechanisms of modified gravity genuinely work in suppressing the fifth force on local scales.  The extendedness of dark matter haloes in radii can provide observables that feature the fifth force, as they exhibit a transient scale of the screening radius.

We have investigated how dark matter haloes evolve in the framework of modified gravity with a screening mechanism, focusing on Vainshtein screening in the Horndeski theory. We have a peculiar interest in
whether the efficiency of the Vainshtein screening relates to the detailed structure of the density profile of a halo.
At first, we revealed that the density profile affects the efficiency of the Vainshtein screening in the inner region of the halo. We found that extended core profiles make the Vainshtein screening less effective in the inner region, even respecting the spherical symmetry of the whole system. By clarifying the efficiency of the Vainshtein screening on the $L$-$c_s$ plot, we postulate that the inner regions where the Vainshtein screening is less effective may keep signatures of the fifth force.
In order to verify our postulate, we provided numerical simulations to trace the dynamical evolution of the halo to equilibrium. We found that the consistent dynamical evolution sustains our postulate. We diagnosed the multiple mass proxies in observations. 

In our simulation, we have treated the fifth force as a central force, approximating the two-body interaction and the long-range asymmetric tidal fields by Newtonian gravity. Of course, this is merely an approximate method and does not accurately trace halo dynamics in the Horndeski theory.
In fact, anisotropic deviations from 
spherical symmetry can make the Vainshtein screening less effective \cite{2013PhRvD..88b1502K, 2019MPLA...3450013O, 2015PhRvD..91h3510B}. We would justify our implementation in the following two points. First, we are interested in whether the Vainshtein screening is efficient or not, even in the set-up in which the Vainshtein screening is stable~\cite{2013PhRvD..88b1502K} inside a core, focusing on the core structure formation through the dynamical evolution of a halo.  Our findings imply that the Vainshtein screening can be less effective even though we respect the spherical symmetry due to the density profile of the halo. Remarkably, the biases in the mass proxies remain even in the case of the NFW profile, where the inner density profile is cuspy with the power $\rho_{\rm dm}(r) \propto r^{-1}$. 
Second, relaxing the spherical symmetry may leave the fifth force unscreened \citep{2013PhRvD..88b1502K,2015PhRvD..91h3510B,2019MPLA...3450013O}. Recent perturbative analyses beyond spherical symmetry \cite{kobayashi_takadera2026} have shown that higher multipoles of the fifth force can remain unscreened even when the monopole is efficiently Vainshtein screened. Therefore, relaxing spherical symmetry does not generically restore screening and may introduce additional modified-gravity signatures.
Our spherical analysis 
should be treated as a leading-order prescription of the unscreened effect of the Vainshtein screening.

Our findings open opportunities to test the Vainshtein screening in the Horndeski theory with the halo in the forthcoming observations of galaxy clusters with multiple wavelengths (see the latest review in \cite{2025arXiv250507697M}). For testing the Vainshtein screening in the Horndeski theory with real data, it is necessary to seriously consider the biases of the mass proxies, which in practice must be more complex. We shall leave the calibration and error estimation issues for 
future study.

\begin{acknowledgements}
We thank T.~Takadera and S.~Saga for useful discussions. This work is partly supported by the Japan Society for the Promotion of Science
(JSPS) Grants-in-Aid for Scientific Research (KAKENHI) Grant No. JP24K17045 (S.A.), JP22K03627, JP23K25868,
JP25K21670 (D.Y.), JP23H00108, JP24K00627~(S.Y.), JP22K03644 (S.M.).
\end{acknowledgements}

\section{Data Availability}
All the mathematical equations are described in the paper, citing appropriate previous studies. The output data of the numerical simulations that are used to plot the results can be openly accessed if there is any request.


\appendix

\section{The definition of model parameters in the Horndeski theory}\label{app:Horndeski_halo_model_detailed}

We review NKYS2013~\cite{Narikawa2013}, which derives the halo model from the field equations in the Horndeski theory. 
The actions of the Horndeski theory and the matter are defined as~\cite{Horndeski1974, Deffayet2009, Kobayashi2011}
\begin{equation}
    S = S_{\rm GG}[g_{\mu\nu},\phi] + S_{\rm m}[g_{\mu\nu},\psi_m]\,.
\end{equation}
with
\begin{equation}
    S_{\rm GG} = \int{\dd^4x \sqrt{-g} (\mathcal{L}_2 + \mathcal{L}_3 + \mathcal{L}_4 + \mathcal{L}_5)}\,,
\end{equation}
where we define
\begin{align}
\mathcal{L}_2 &= K(\phi,X)\,, \cr
\mathcal{L}_3 &= - G_{3}(\phi,X)\Box\phi\,,\cr
\mathcal{L}_4 &= G_{4}(\phi,X)R + G_{4X}\big[(\Box \phi)^2 - (\nabla_{\mu}\nabla_{\nu}\phi)^2 \big]\,, \cr
\mathcal{L}_5 &= G_{5}(\phi,X)G_{\mu \nu}\nabla^{\mu}\nabla^{\nu}\phi \cr
& - \frac16G_{5X}\big[(\Box\phi)^3 - 3\Box\phi(\nabla_\mu \nabla_\nu\phi)^2 + 2(\nabla_\mu \nabla_\nu\phi)^3 \big]\,, \cr
\label{eq:Horndeski_Lagrangean}
\end{align}
Here $G_{iX}\equiv \partial G_i/\partial X$ and $X$ denote the canonical kinetic energy of the scalar field as $X \equiv -\frac12 g^{\mu\nu}\nabla_\mu\phi\,\nabla_\nu\phi$. 
In general, a scalar field $\phi$ evolves on cosmological timescales and may drive cosmic expansion. However, our interest lies in gravitational phenomena on scales much smaller than the Hubble radius, such as galaxy clusters and galactic halos. 
In this sub-horizon regime, the effect of cosmic expansion can be treated as a higher-order correction, and a local patch approximation is valid.

It is known that the time variation of the background scalar field $\phi_0(t)$ may induce a time variation of the effective gravitational constant. 
However, Lunar Laser Ranging experiments place strong constraints on such variation. Therefore, we assume that over the timescale of interest the background scalar field is effectively constant:
\begin{align}
\phi_0 \simeq \mathrm{const.}, \qquad X_0 \simeq 0.
\end{align}

Accordingly, we adopt the Minkowski metric as the background spacetime and take the scalar field to be constant:
\begin{align}
\mathrm{d}s^2 &= \eta_{\mu\nu}\mathrm{d}x^\mu \mathrm{d}x^\nu,\\
\phi &= \phi_0 = \mathrm{const.}, \quad X_0 = 0.
\end{align}

Throughout this section, the functions $K, G_i$ and their derivatives with respect to $\phi$ and $X$ (e.g., $G_{4X}$) are evaluated at the background values $(\phi_0, X_0)$ and treated as constants unless otherwise stated.

To allow this background solution, we require
\begin{align}
K(\phi_0,0)=0, \qquad K_\phi(\phi_0,0)=0,
\end{align}
where the former eliminates an effective cosmological constant, and the latter ensures the background scalar-field equation is satisfied.

Following the derivation in Ref.~\cite{Narikawa2013}, the arbitrary functions of the Horndeski theory are reparametrized into
\begin{align}
G_4 &= \frac{M_{\rm Pl}^2}{2}, \cr
\qquad
G_{4\phi} &= M_{\rm Pl}\xi,
\cr
K_X-2G_{3\phi} &= \eta, \notag\\
-G_{3X}+3G_{4\phi X} &= \frac{\mu}{\Lambda^3}, \notag\\
G_{4X}-G_{5\phi} &= \frac{M_{\rm Pl}}{\Lambda^3}\alpha, \notag\\
G_{4XX}-\frac{2}{3}G_{5\phi X} &= \frac{\nu}{\Lambda^6}, \notag\\
G_{5X} &= -\frac{3M_{\rm Pl}}{\Lambda^6}\beta.
\label{eq:parametrize}
\end{align}

This corresponds, for example, to expanding the arbitrary functions around $(\phi_0,X_0)$ as

\begin{align}
K(\phi,X) &\approx \eta X, \notag\\
G_3(\phi,X) &\approx -\frac{\mu}{\Lambda^3}X, \notag\\
G_4(\phi,X) &\approx \frac{M_{\rm Pl}^2}{2}
+M_{\rm Pl}\xi\phi
+M_{\rm Pl}\frac{X}{\Lambda^3}\alpha
+\frac12\left(\frac{X}{\Lambda^3}\right)^2\nu,
\notag\\
G_5(\phi,X) &\approx -\frac{M_{\rm Pl}X}{\Lambda^6}\beta.
\label{eq:parametrized_function}
\end{align}

Although the expansion itself is not unique, if the combinations of derivatives in Eq.~\eqref{eq:parametrize} coincide, the quasi-static equations take the same form.
%
%
\section{Dictionary of Halo Profiles}
\label{app:halo_profiles}

For a spherically symmetric density profile $\rho(r)$, we introduce a
model-dependent characteristic radius $r_\star$ and density
$\rho_\star$, and define
\begin{equation}
x:=\frac{r}{r_\star},
\qquad
g(x):=\frac{\rho(r_\star x)}{\rho_\star}.
\end{equation}
The enclosed mass can then be written as
\begin{equation}
M(<r)
=
\int_0^r \D r'\,4\pi r'^2\rho(r')
=
4\pi\rho_\star r_\star^3 f_m(x),
\end{equation}
where the dimensionless mass shape factor is
\begin{equation}
f_m(x)
:=
\int_0^x \D u\,u^2g(u).
\end{equation}

The dimensionless density and enclosed-mass profiles for the halo
models considered in this work are summarized in
Table~\ref{tab:halo_profile_dictionary}.

\begin{table*}[t]
\centering
\caption{
Dictionary of the dimensionless halo profiles used in this work.
For each model, the density and enclosed mass are written as
$\rho(r)=\rho_\star g(x)$ and
$M(<r)=4\pi\rho_\star r_\star^3f_m(x)$, respectively, where
$x=r/r_\star$.
}
\label{tab:halo_profile_dictionary}

\renewcommand{\arraystretch}{2.2}
\setlength{\tabcolsep}{3pt}

\begin{tabular}{lccc}
\hline
Profile
&
Characteristic scales
&
Dimensionless density $g(x)$
&
Dimensionless mass factor $f_m(x)$
\\
\hline
\hline

NFW
&
$\begin{gathered}
r_\star=r_s,\\
\rho_\star=\rho_s
\end{gathered}$
&
$\displaystyle
\frac{1}{x(1+x)^2}
$
&
$\displaystyle
\ln(1+x)-\frac{x}{1+x}
$
\\
\hline

Pseudo-isothermal
&
$\begin{gathered}
r_\star=r_{\rm core},\\
\rho_\star=\rho_{\rm core}
\end{gathered}$
&
$\displaystyle
\frac{1}{1+x^2}
$
&
$\displaystyle
x-\arctan x
$
\\
\hline

Burkert
&
$\begin{gathered}
r_\star=r_{\rm core},\\
\rho_\star=\rho_{\rm core}
\end{gathered}$
&
$\displaystyle
\frac{1}{(1+x)(1+x^2)}
$
&
$\displaystyle
\frac{1}{2}\ln(1+x)
+\frac{1}{4}\ln(1+x^2)
-\frac{1}{2}\arctan x
$
\\
\hline

\end{tabular}
\end{table*}

For the NFW profile, the characteristic density can alternatively be
expressed in terms of $M_{200}$ and the concentration $c_{200}$.
Defining
\begin{equation}
R_{200}=c_{200}r_s,
\qquad
M_{200}
=
\frac{4\pi}{3}
\left(200\rho_{\rm ref}\right)R_{200}^3,
\end{equation}
and requiring $M(<R_{200})=M_{200}$ gives
\begin{equation}
\rho_s
=
\frac{200}{3}\rho_{\rm ref}
\frac{c_{200}^3}
{\ln(1+c_{200})-\dfrac{c_{200}}{1+c_{200}}}.
\end{equation}
Here, $\rho_{\rm ref}$ is either the critical density
$\rho_{\rm crit}$ or the mean matter density, depending on the adopted
halo-mass convention.
\section{Weak lensing mass estimation based on $N$-body simulations}
\label{app:lensing}

Here we discuss our numerical implementation for weak lensing convergence.
First, the surface density associated with the Newtonian contribution is
constructed by projecting the particle masses along the line of sight as
\begin{align}
\sum_i m_i \rightarrow \Sigma_{\rm N}(R)\,.
\end{align}
The corresponding convergence is then given by
$\kappa_{\rm N}=\Sigma_{\rm N}/\Sigma_{\rm crit}$.
For the contribution associated with the fifth force in the convergence, $\kappa_{5}$, 
we use the scalar-field quantity
$x(r)$ stored in the snapshot to evaluate
\begin{align}
\left[(\alpha x^2+2\beta x^3)r^3\right]'\,,
\end{align}
by numerical differentiation with respect to the particle radius.
We then project this quantity over the finite interval
$Z\in[-Z_{\max},Z_{\max}]$, and construct $\kappa_{5}$.
Because the central factor $1/r^2$ is susceptible to numerical
divergence as $r\to0$, we introduce a regularized radius $r_{\rm safe}$
in the numerical implementation.

As for the shear catalog with the shape noise, we first calculate
\begin{align}
\gamma_t(R)=\bar{\kappa}(<R)-\kappa(R)\,,
\end{align}
from $\kappa(R)$ using a cumulative sum over discrete radial bins. We then distribute background galaxies uniformly over the projected
plane, assign the true tangential shear $\gamma_t(R)$, and add shape
noise.
The synthetic catalog is generated according to
\begin{align}
g_1&=-\gamma_t(R)\cos(2\phi)+\mathcal{N}(0,\sigma_e)\,, \cr
g_2&=-\gamma_t(R)\sin(2\phi)+\mathcal{N}(0,\sigma_e)\,,
\end{align}
where $\mathcal{N}(x,\sigma_e)$ represents the Gaussian shape noise with the mean $x$ and the variance $\sigma_e$. 
Here, we assumed $x=0$.
We construct the azimuthally averaged shear $\hat{\gamma}_t(R)$ by averaging over annular bins.
The uncertainty in each bin is estimated as the standard error of the
mean derived from the within-bin variance and therefore scales as
$1/\sqrt{N_b}$.
We fit $\hat{\gamma}_t(R)$ to the NFW and Burkert shear models
using weighted least squares.
Here, $\Sigma_{\rm crit}$ is treated as a constant and absorbed into the
units, so that $\gamma_t\simeq\Delta\Sigma$.
The three-dimensional mass profile $M_{\rm WL}(<r)$ is then
reconstructed from the maximum-likelihood parameter.

Note that the weak-lensing mass estimate relies on fitting with the model of the halo density profile.
If MG
were to produce a significant deformation of
the halo density profile, this model mismatch could introduce a
systematic error in $M_{\rm WL}$.
As a representative test, we checked the validity of the NFW model even in the MG using generalized least squares.

We confirmed that the NFW model provides a generally good description
over the radial range primarily considered in this chapter.
Thus, within the setup adopted here, 
MG does not strongly
distort the halo density profile, and the use of the NFW model for the
weak-lensing fit is justified.

The goodness of fit can nevertheless depend on whether the outermost
radial bins are included, or equivalently, on the adopted value of
$r_{\max}$.
This analysis should therefore be regarded as a diagnostic of the
validity of the weak-lensing fitting model rather than as a precise test
of the underlying theory of gravity.

\section{Sanity checks for numerical accuracy}
\label{sec:numerical_sanity}

This appendix provides details of our numerical computation for readers who may
wish to assess the numerical accuracy of the simulations.
We quantitatively test how accurately our numerical scheme computes the
fifth-force response of the $N$-body particles by examining the initial
force solution, the temporal evolution of the energy budget, and the
stability of the equilibrium density profile.
We confirm that (1) the total-energy budget fluctuates only at the level of
$\sim 10^{-4}$, and (2) the acceleration induced by the fifth force
is accurately recovered over a wide range of radial scales.

\begin{figure*}[t]
    \centering
    \includegraphics[width=0.92\linewidth]{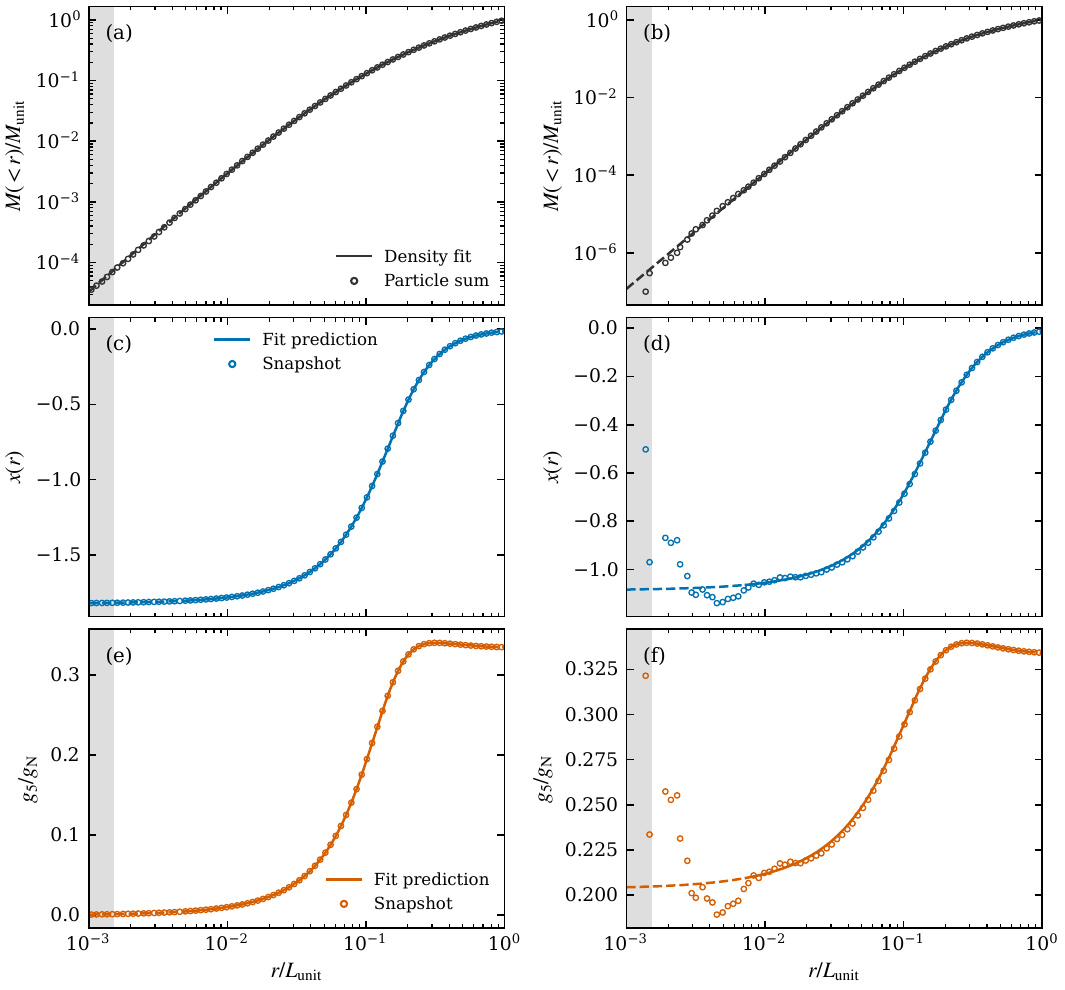}
    \caption{
    Initial-condition and force-solver sanity check.
    The left and right columns show the two benchmark configurations adopted
    for the numerical test.
    Panels (a) and (b) show the enclosed mass $M(<r)$:
    the black solid curves denote the mass obtained by integrating the fitted
    smooth density profile (``Density fit''), while the open circles denote
    the direct particle sum (``Particle sum'').
    Panels (c) and (d) compare the radial field variable $x(r)$.
    The blue solid curves show the prediction obtained from the fitted density
    profile (``Fit prediction''), and the blue open circles show the values
    measured from the $N$-body snapshot (``Snapshot'').
    Panels (e) and (f) show the corresponding ratio of the fifth-force and
    Newtonian accelerations, $g_5/g_{\rm N}$.
    Here the orange solid curves are the fit-based predictions and the orange
    open circles are the snapshot measurements.
    Agreement between the curves and symbols over most of the radial range
    demonstrates that the particle realization and the fifth-force solver
    consistently reproduce the smooth-profile prediction; the enhanced
    scatter at the smallest radii is caused by the small number of particles
    in the innermost shells.
    }
    \label{fig:q5_initial}
\end{figure*}

We first test whether the particle realization and the force solver reproduce
the profiles expected from the smooth input density distribution.
Figure~\ref{fig:q5_initial} compares the cumulative mass profile obtained
directly by summing the particles with that inferred from the fitted density
profile.  The two estimates are almost indistinguishable over the resolved
radial range in both benchmark configurations, demonstrating that the
particle sampling accurately represents the assumed smooth halo profile.
The larger scatter at the smallest radii is expected because only a small
number of particles contribute to the innermost shells.

The middle and bottom rows of Fig.~\ref{fig:q5_initial} provide a more direct
test of the fifth-force calculation.
The radial solution $x(r)$ obtained from an individual particle snapshot
closely follows the prediction evaluated from the smooth density fit.
The resulting acceleration ratio, $g_5/g_{\rm N}$, likewise agrees with the
smooth-profile prediction over several decades in radius.
The fifth-force response changes appreciably with radius---from a strongly
suppressed value in the inner region to a finite fraction of the Newtonian
acceleration at intermediate and large radii---and this radial transition is
reproduced by the particle calculation.
The visible scatter in the innermost points, especially in the right-hand
configuration, coincides with the region in which the enclosed particle
number becomes small.  We therefore regard it as particle-sampling noise
rather than a systematic failure of the fifth-force solver.

\begin{figure}[t]
    \centering
    \includegraphics[width=0.78\linewidth]{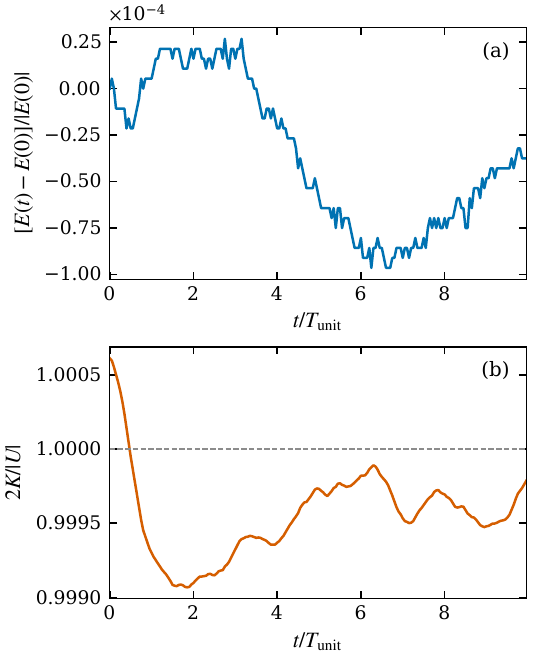}
    \caption{
    Global conservation test of the $N$-body integration.
    Panel (a) shows the fractional total-energy error,
    $[E(t)-E(0)]/|E(0)|$, as a function of time in units of
    $T_{\rm unit}$.  The blue solid curve remains bounded at approximately
    the $10^{-4}$ level throughout the calculation.
    Panel (b) shows the virial ratio $2K/|U|$ as the orange solid curve.
    The horizontal gray dashed line marks the equilibrium value
    $2K/|U|=1$.
    }
    \label{fig:q4_energy_conserve}
\end{figure}

A complementary test is provided by the temporal evolution of the global
energy budget.
Figure~\ref{fig:q4_energy_conserve}(a) shows the fractional change in the
total energy relative to its initial value.
Throughout the simulated interval, the quantity
$[E(t)-E(0)]/|E(0)|$ remains within approximately $10^{-4}$.
There is a slow, bounded excursion rather than a secular growth of the
energy error, indicating that the accumulated integration error remains
small over the duration relevant to our analysis.

Figure~\ref{fig:q4_energy_conserve}(b) shows the virial ratio
$2K/|U|$.
After a small initial transient, the system remains close to the equilibrium
value of unity.
The residual variation is at the $\sim10^{-3}$ level and is much smaller
than the changes induced by the fifth-force dynamics discussed below.
The simultaneous stability of the energy budget and of the virial ratio
therefore indicates that the subsequent evolution is not driven by a
spurious numerical heating or cooling of the particle system.

\begin{figure}[t]
    \centering
    \includegraphics[width=0.78\linewidth]{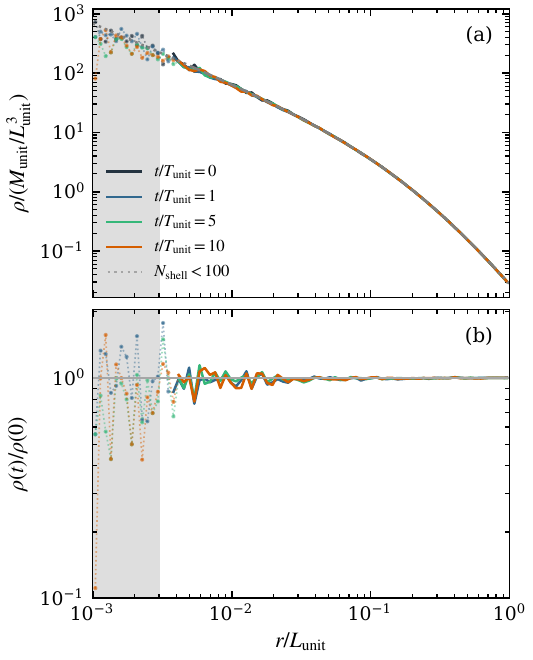}
    \caption{
    Stability of the equilibrium halo under Newtonian evolution.
    Panel (a) shows the density profile $\rho(r)$ at
    $t/T_{\rm unit}=0$ (black), $1$ (blue), $5$ (green), and $10$
    (orange).
    The profiles overlap over essentially the entire resolved radial range.
    Panel (b) shows the same evolution as the ratio
    $\rho(t)/\rho(0)$; unity therefore corresponds to an exactly stationary
    density profile.
    The gray shaded region at small radius denotes shells containing fewer
    than $100$ particles ($N_{\rm shell}<100$), where the density estimate
    is dominated by particle shot noise.
    The fluctuations confined to this shaded region should consequently not
    be interpreted as physical evolution.
    Outside it, the density ratio remains close to unity, demonstrating the
    numerical stability of the Newtonian equilibrium configuration.
    }
    \label{fig:density_evolution_newton}
\end{figure}

We next test whether the initial halo remains stationary when evolved with
Newtonian gravity alone.
This provides an especially useful control experiment because an equilibrium
initial condition should exhibit no systematic change in its density profile
in the absence of the additional force.
Figure~\ref{fig:density_evolution_newton}(a) compares the density profiles at
$t/T_{\rm unit}=0$, $1$, $5$, and $10$.
Outside the innermost particle-poor region, the profiles at all four epochs
lie almost exactly on top of one another.
The same conclusion is more apparent in
Fig.~\ref{fig:density_evolution_newton}(b), where the density is divided by
its initial value.
The ratio remains very close to unity throughout the resolved radial range.

At very small radii the ratio becomes noisy.
This region contains fewer than $100$ particles per radial shell and is
therefore dominated by discreteness noise.
Importantly, no coherent outward propagation or secular deformation of the
density profile is observed outside this low-particle-number region.
The Newtonian control run therefore demonstrates that the initial profile
is sufficiently well equilibrated and that numerical relaxation does not
produce an appreciable artificial evolution on the timescales considered
here.

\begin{figure*}[t]
    \centering
    \includegraphics[width=0.92\linewidth]{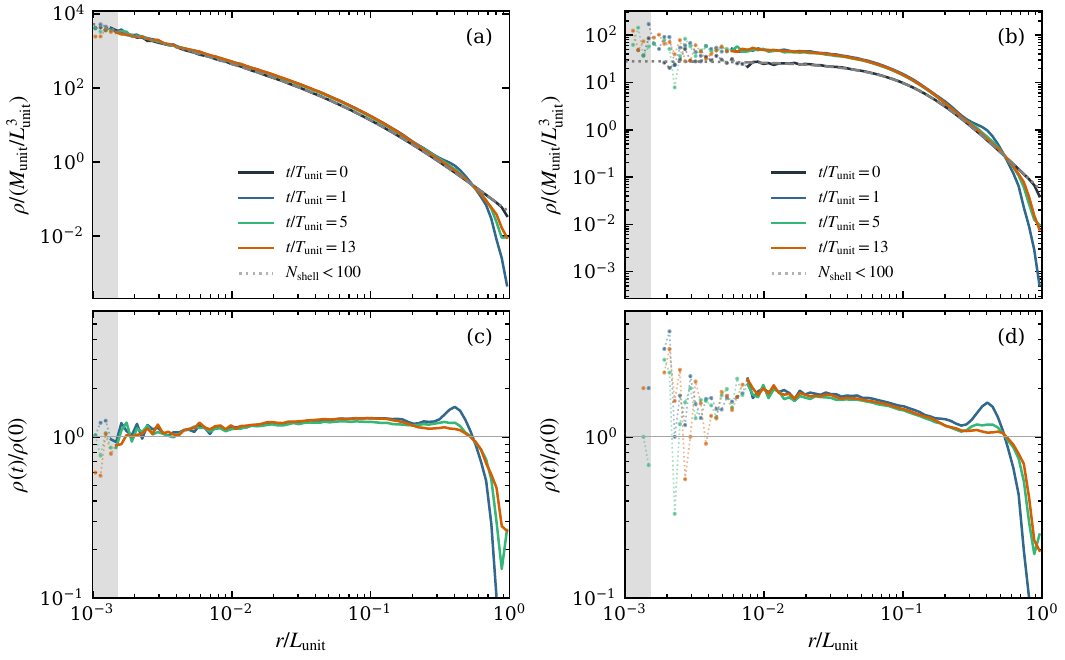}
    \caption{
    Evolution of the density profile in the presence of the modified-gravity
    force.
    The left and right columns correspond to the two benchmark
    modified-gravity configurations used in the numerical experiment.
    Panels (a) and (b) show $\rho(r)$ at
    $t/T_{\rm unit}=0$ (black), $1$ (blue), $5$ (green), and $13$
    (orange).
    Panels (c) and (d) show the corresponding ratios
    $\rho(t)/\rho(0)$, for which unity represents the initial density
    distribution.
    In all panels, the gray shaded region marks radial shells containing
    fewer than $100$ particles ($N_{\rm shell}<100$); fluctuations within
    this region are therefore dominated by discreteness noise.
    Outside the shaded region the later-time curves display a smooth,
    coherent departure from the initial profile.
    This behavior differs qualitatively from the nearly stationary Newtonian
    control run in Fig.~\ref{fig:density_evolution_newton}, supporting the
    interpretation that the density evolution is generated by the physical
    fifth-force response rather than by numerical relaxation.
    }
    \label{fig:density_evolution_modigrav}
\end{figure*}

Finally, Fig.~\ref{fig:density_evolution_modigrav} repeats the density-profile
comparison after the modified-gravity force is switched on.
In contrast to the Newtonian control run, a coherent and radially correlated
change in the density distribution develops with time.
The upper panels show that the later-time profiles depart systematically from
the initial profile over radii containing a large number of particles.
The corresponding ratios in the lower panels make this evolution explicit:
$\rho(t)/\rho(0)$ develops a broad deviation from unity rather than the
uncorrelated point-to-point scatter seen in the poorly sampled central bins.

The contrast between Figs.~\ref{fig:density_evolution_newton} and
\ref{fig:density_evolution_modigrav} is an important numerical consistency
check.
The Newtonian system preserves the initial equilibrium configuration, whereas
the modified-gravity simulations exhibit a systematic response only after
the additional force is introduced.
Moreover, this response occurs predominantly outside the region flagged by
the $N_{\rm shell}<100$ criterion.
Together with the direct force comparison in Fig.~\ref{fig:q5_initial} and
the energy-conservation test in Fig.~\ref{fig:q4_energy_conserve}, these
results indicate that the measured redistribution of particles is a physical
response to the fifth force rather than an artifact of particle sampling or
time integration.

\clearpage
\bibliographystyle{apsrev4-1}
\onecolumngrid
\bibliography{myref}

\end{document}